\documentclass[%
 aip,
 amsmath,amssymb,
 reprint,
]{revtex4-1}

\usepackage{soul}
\usepackage{graphicx}

\usepackage{dcolumn}
\newcolumntype{z}[1]{D{.}{.}{#1}}

\usepackage{bm}
\usepackage{MnSymbol}
\usepackage{siunitx}

\newcommand{\e}{\mathrm{e}}
\newcommand{\diff}{\mathrm{d}}
\newcommand{\kB}{k_\mathrm{B}}
\newcommand{\mbr}{\mathbf{r}}

\newcommand{\mbE}{\mathbf{E}}
\newcommand{\mbm}{\mathbf{m}}

\newcommand{\mbX}{\mathbf{X}}
\newcommand{\mbpi}{\boldsymbol{\pi}}
\newcommand{\mbalpha}{\boldsymbol{\alpha}}

\begin{document}

\title[Third dielectric virial coefficient of helium]{Path-integral
  calculation of the third dielectric virial coefficient of helium based on
  {\em ab initio} three-body polarizability and dipole surfaces} 

\author{Giovanni Garberoglio}
\email[Corresponding author: ]{garberoglio@ectstar.eu}
\affiliation{European Centre for Theoretical Studies in Nuclear Physics and
  Related Areas (FBK-ECT*), Trento, I-38123, Italy.}
\affiliation{Trento Institute for Fundamental Physics and Applications
  (INFN-TIFPA), via Sommarive 14, 38123 Trento, Italy}

\author{Allan H. Harvey}
\email{allan.harvey@nist.gov}
\affiliation{Applied Chemicals and Materials Division, National Institute of
  Standards and Technology, Boulder, CO 80305, USA.}

\author{Jakub Lang}
\email{jakub.lang@chem.uw.edu.pl}
\affiliation{Faculty of Chemistry, University of Warsaw, Pasteura 1, 02-093
  Warsaw, Poland.}

\author{Micha\l{} Przybytek}
\email{m.przybytek@uw.edu.pl}
\affiliation{Faculty of Chemistry, University of Warsaw, Pasteura 1, 02-093
  Warsaw, Poland.}

\author{Micha\l{} Lesiuk}
\email{m.lesiuk@uw.edu.pl}
\affiliation{Faculty of Chemistry, University of Warsaw, Pasteura 1, 02-093
  Warsaw, Poland.}

\author{Bogumi\l{} Jeziorski}
\altaffiliation{Deceased}
\affiliation{Faculty of Chemistry, University of Warsaw, Pasteura 1, 02-093
  Warsaw, Poland.}

\date{\today}

\begin{abstract}
We develop a surface for the electric dipole moment of three interacting helium atoms and
use it, together with state-of-the-art potential and polarizability surfaces, to compute the third
dielectric virial coefficient, $C_\varepsilon$, for both $^4$He and $^3$He isotopes.  Our results
agree with previously published data computed using an approximated form for the three-body
polarizability, and are extended to the low-temperature regime by including exchange effects.
Additionally, the uncertainty of $C_\varepsilon$ is rigorously determined for the first time by
propagating the uncertainties of the potential and polarizability surfaces; this uncertainty is much
larger than the contribution from the dipole-moment surface to $C_\varepsilon$.  Our results compare
reasonably well with the limited experimental data. The first-principles values of $C_\epsilon$
computed in this work will enhance the accuracy of primary temperature and pressure metrology based
on measurements of the dielectric constant of helium.
\end{abstract}

\maketitle

\section{Introduction}

Primary metrology for pressure and thermodynamic temperature has been
transformed in recent years by the ability to make precise electrical,
optical, and acoustic measurements on noble gases.
Examples include dielectric-constant gas
thermometry,\cite{Gaiser_2015,Gaiser_2017,Gaiser_2020a} refractive-index
gas thermometry,\cite{Rourke_2019,Gao_2020,Ripa_2021,Rourke_2021} acoustic
gas thermometry,\cite{Moldover_2014} and \textit{ab initio} pressure
standards based on dielectric measurements.\cite{Gaiser20,Gaiser_2022}

The use of helium gas for such standards is particularly attractive,
because the small number of electrons makes it possible to compute
properties of atoms, pairs of atoms, and to some extent triplets of atoms
with extremely small uncertainties, much smaller than can be obtained from
experiment.~\cite{Review2023} An example is the static polarizability of
the $^4$He atom, which has been computed with a relative standard
uncertainty near $1 \times 10^{-7}$ (0.1 ppm).\cite{Puchalski20}

For dielectric-based standards, the dielectric virial coefficients are essential quantities. These coefficients appear in the expansion around the zero-density limit of the Clausius--Mossotti function in powers of molar density $\rho$:
\begin{equation}
  \frac{\varepsilon - 1}{\varepsilon + 2} = \rho \left( A_\varepsilon + B_\varepsilon\rho
  + C_\varepsilon\rho^2 + \ldots \right) ,
\label{eq:Vir:eps}
\end{equation}
where $\varepsilon$ is the static dielectric constant and $A_\varepsilon$ is proportional to the mean static polarizability of the isolated molecule.
$B_\varepsilon$ is the second dielectric virial coefficient, $C_\varepsilon$ is the third dielectric
virial coefficient, etc. These virial coefficients are functions of the thermodynamic temperature $T$.

The second dielectric virial coefficient $B_\varepsilon(T)$ depends only on the pair potential and
the interaction-induced pair polarizability; for helium the former has been calculated with
extremely high accuracy\cite{u2_2020} and the latter is also fairly well known.\cite{Cencek11}
Garberoglio and Harvey computed $B_\varepsilon$ of helium, along with the related second
refractivity virial coefficient, with full accounting for quantum effects.\cite{Garberoglio20:Beps}
The expanded ($k = 2$) uncertainty in $B_\varepsilon$ was only \SI{0.2}{cm^6.mol^{-2}} at most
temperatures, increasing to \SI{0.4}{cm^6.mol^{-2}} as the temperature decreased to 1~K.  At
temperatures near 273~K, this is less than 0.5\% in relative terms.
  
Calculation of the third dielectric virial coefficient $C_\varepsilon(T)$ is much more complex.
Recently, three of us\cite{Garberoglio_2021} presented the first complete framework for calculation
of $C_\varepsilon$, making use of a path-integral Monte Carlo (PIMC) method to fully account for
quantum effects.  In addition to the pair quantities required for $B_\varepsilon$, this calculation
requires multidimensional surfaces for the three-body non-additive potential energy, for the
three-body non-additive polarizability, and for the dipole moment of a three-atom configuration.
Because surfaces were not available for the three-body polarizability and dipole moment, Garberoglio
\textit{et al.}\cite{Garberoglio_2021} approximated them with simple expressions valid in the limit
of large interatomic distances, so their calculated results were not rigorous and lacked uncertainty
estimates. In the meantime, a fully {\em ab initio} three-body polarizability surface has been
developed~\cite{a3_2022} and the non-additive three-body potential~\cite{u3_2023} has been improved.

In this work, we present the dipole-moment surface of an assembly of three helium atoms, provide a
brief review of the theoretical framework for calculating $C_\varepsilon$, and use the developed
dipole-moment surface, together with the latest three-body potential and polarizability, to compute
$C_\varepsilon$, with uncertainty estimates, from $1$~K to $3000$~K.  We compare our calculations to
the limited (and scattered) experimental data available for this quantity, noting that the
uncertainties in our calculated $C_\varepsilon(T)$ are much smaller than those of the experimental
data.

\section{Three-body dipole surface}
\label{sec:dipole}

For three interacting atoms located at fixed positions in space specified by vectors $\mathbf{r}_i$,
$i\in\{1,2,3\}$, the electric dipole moment $\mbm(\mathbf{r}_1,\mathbf{r}_2,\mathbf{r}_3)$ is
defined as a vector
\begin{equation}\label{polar_definition}
m_{\alpha}(\mathbf{r}_1,\mathbf{r}_2,\mathbf{r}_3) = - 
\frac{\partial \mathcal{E}(\mathbf{r}_1,\mathbf{r}_2,\mathbf{r}_3,\mathbf{E})}{\partial E_\alpha}
  \bigg|_{ \mathbf{E}=0},
\end{equation}
where $\alpha \in \{x,y,z\}$ and 
$\mathcal{E}(\mathbf{r}_1,\mathbf{r}_2,\mathbf{r}_3,\mathbf{E})$ is the
electronic energy in the presence of a static and uniform electric field
$\mathbf{E}$ with components $E_\alpha$. 
As the dipole moment of atoms and homonuclear diatoms vanishes, 
$\mbm(\mathbf{r}_1,\mathbf{r}_2,\mathbf{r}_3)$ for a system of three interacting helium atoms 
is effectively equal to its three-body component $\mbm_3(\mathbf{r}_1,\mathbf{r}_2,\mathbf{r}_3)$.
This observation allows us to remove the basis set superposition error (BSSE)~\cite{Boys1970} 
from the calculated dipole moments using the complete counterpoise correction,\cite{Gutowski:86}
which has been shown to be applicable not only in the calculations of interaction energies\cite{Boys1970,Gutowski:86,Chalasinski:00}
but also for interaction-induced dipole moments and polarizabilities.~\cite{2009TChA..122...127H,a3_2022}
The counterpoise corrected three-body dipole moment 
$\mbm_3 (\mathbf{r}_1,\mathbf{r}_2,\mathbf{r}_3)$ is obtained by combining results 
from seven separate calculations,
\begin{equation}\label{dipole_total}
\begin{split}
\mbm_3 &(\mathbf{r}_1,\mathbf{r}_2,\mathbf{r}_3 )=
\mbm(\mathbf{r}_1,\mathbf{r}_2,\mathbf{r}_3)_B \\
&- \mbm(\mathbf{r}_1,\mathbf{r}_2)_B
 - \mbm(\mathbf{r}_2,\mathbf{r}_3)_B 
 - \mbm(\mathbf{r}_1,\mathbf{r}_3)_B \\
&+ \mbm(\mathbf{r}_1)_B
 + \mbm(\mathbf{r}_2)_B
 + \mbm(\mathbf{r}_3)_B,
\end{split}
\end{equation}
where $\mbm(\mathbf{r}_1,\mathbf{r}_2,\mathbf{r}_3)_B$ is the dipole moment of the full trimer,
$\mbm(\mathbf{r}_i,\mathbf{r}_j)_B$ is the dipole moment of the diatom 
comprising atoms located at $\mathbf{r}_i$ and $\mathbf{r}_j$, 
and $\mbm(\mathbf{r}_i)_B$ is the dipole moment of the $i$-th atom. All contributions
on the right-hand side of Eq.~(\ref{dipole_total}) are computed using the same
basis set $B$ of the trimer. 
Note that, due to the presence of basis functions located on all three atomic centers, 
the diatomic dipole moments $\mbm(\mathbf{r}_i,\mathbf{r}_j)_B$ 
are not necessarily zero and depend on the position of the third atom. 
Analogously, the atomic dipole moments $\mbm(\mathbf{r}_i)_B$ depend on 
the position of the other two atoms. The contributions $\mbm(\mathbf{r}_i,\mathbf{r}_j)_B$ and $\mbm(\mathbf{r}_i)_B$ vanish only in the complete basis set limit.

Before delving into computational details, we first analyze the accuracy requirements in calculation
of the three-body dipole moment of helium from the point of view of the present work. Note that the
contribution of the three-body dipole moment to the third dielectric virial coefficient is expected
to be small in comparison with the three-body polarizability, especially for higher temperatures. In
fact, the three-body dipole moment enters the calculations through the factor
$|\mbm_3|^2/(k_\mathrm{B}T)$ and, additionally, the square of the three-body dipole moment,
$|\mbm_3|^2$, is smaller by one to two orders of magnitude than the corresponding value of the
isotropic three-body polarizability (when both quantities are expressed in atomic units). Taking
into consideration that the average uncertainty in the best theoretical results currently available
in the literature for the three-body polarizability is of the order of several percent,
uncertainties larger by a factor of ten in the three-body dipole moment are still acceptable.

We performed \emph{ab initio} calculations of the three-body dipole moment of helium 
for 734 atomic configurations defined in terms of interatomic distances 
$r_{12}$, $r_{23}$, and $r_{13}$ that satisfy the triangle inequality. 
This set contains a subset of configurations used in Ref.~\onlinecite{a3_2022} (600 points),
obtained by excluding equilateral and centrosymmetric linear configurations 
for which the dipole moment vanishes by symmetry.
The additional points were generated using the incremental procedure described 
in Sec.~III.B of Ref.~\onlinecite{a3_2022} and Sec.~3.2 of Ref.~\onlinecite{u3_2023}.
If necessary, the interatomic distances defining initial configurations were relabeled
to ensure that they are sorted according to their magnitude, that is, 
$r_{12} \leq r_{23} \leq r_{13}$.
In this case, the angle $\theta_2$ at the vertex corresponding to atom 2, 
which can be calculated from the law of cosines as
\begin{equation}
\theta_2=\arccos\frac{r_{12}^2+r_{23}^2-r_{13}^2}{2r_{12}r_{23}},
\end{equation}
necessarily satisfies the condition $\theta_2 \geq \pi/3$.
During the calculations, the atoms were assumed to lie in the \textit{xy}-plane 
with atoms $1$ and $2$ placed along the \textit{x}-axis.
With this choice, the $z$ component of the dipole moment is equal to zero for all configurations.

\begin{table*}
\caption{Basis set convergence of the length of the three-body dipole moment $|\mbm_3|$ of helium calculated at the CC3 level of theory using the d$X$Z family of basis sets. 
The interatomic distances, $r_{ij}$, are in units of $a_0$ and the electric dipole moment is in units of $10^{-3}~e~a_0 \approx 8.478354 \times 10^{-33}$~C~m.}
\label{tab:basis_conv}
\begin{ruledtabular}
\begin{tabular}{z{2.2}z{2.2}z{2.2}z{1.6}z{1.6}z{1.6}z{1.6}z{1.6}}
\multicolumn{1}{c}{$r_{12}$} &
\multicolumn{1}{c}{$r_{23}$} &
\multicolumn{1}{c}{$r_{13}$} & 
\multicolumn{1}{c}{$X = 2$} & 
\multicolumn{1}{c}{$X = 3$} & 
\multicolumn{1}{c}{$X = 4$} &
\multicolumn{1}{c}{$X = 5$} &
\multicolumn{1}{c}{$X = 6$}\\
\hline
   3.01&   7.28&   8.96&    0.066435&    0.064043&    0.063199&    0.062926&    0.062791\\
   3.91&   4.41&   6.76&    0.130613&    0.118156&    0.114890&    0.114201&    0.113865\\
   4.40&   5.88&   8.23&    0.013709&    0.011621&    0.011066&    0.010938&    0.010891\\
   4.51&   5.78&   9.97&    0.012897&    0.012214&    0.011845&    0.011791&    0.011757\\
   3.07&   4.97&   4.97&    0.280883&    0.303940&    0.304889&    0.304283&    0.303943\\
   7.56&  10.94&  12.06&    0.000003&    0.000012&    0.000013&    0.000014&    0.000014\\
   7.56&  12.06&  15.44&    0.000003&    0.000009&    0.000010&    0.000010&    0.000010\\
\end{tabular}
\end{ruledtabular}
\end{table*}

The dipole moments were calculated using the CC3 method \cite{hald2003calculation} as implemented in
the Dalton 2018 package.~\cite{dalton,dalton2018}. We used tight convergence criteria,
  between $10^{-7}$ and $10^{-10}$ depending on the geometry, for the coupled-cluster ground state
  and response functions iterations, as molecular properties tend to be sensitive to such
  thresholds. In general, for smaller interatomic distances it is more difficult to achieve
  convergence and the thresholds were looser in such cases. Nonetheless, at least four significant
  digits are stable in the results for all geometries of the trimer.

The calculations were performed using a family of doubly augmented correlation-consistent Gaussian
basis sets, referred to further as d$X$Z, where $X$ is the cardinal number. These basis sets were
developed in Ref.~\onlinecite{cencek2012effects} specifically to accurately describe interactions of
helium atoms in their ground state. The most important difference between the d$X$Z basis sets and
the standard Dunning's d-aug-cc-pV$X$Z basis sets for helium~\cite{Dunning1989cc,Dunning1994he} is
that in the former the augmenting functions, optimized to accurately reproduce the asymptotic
expansion of the pair potential, are more diffuse (have smaller exponents) than in the latter,
especially for angular momenta $l\ge2$. Therefore, the d$X$Z basis sets are more suitable for
calculations of molecular properties, such as the dipole moment and polarizability, that require
proper description of the electronic wave function at large distances from the nuclei.

The largest basis set of this family which is currently available is d6Z. Unfortunately,
calculations within the d6Z basis are possible only for a handful of configurations due to the
enormous computational costs. Nonetheless, we obtained d6Z results for a few configurations in order
to test the basis set convergence of the calculated data. This allows us to select the basis which
yields sufficient accuracy at minimal possible cost.

In Table~\ref{tab:basis_conv}, we present the results obtained for several representative
configurations.  For a vast majority of configurations, the d4Z values are essentially converged and
are very close to the d5Z and d6Z results.  For example, for the isosceles configuration with side
lengths $(r_{12},r_{23},r_{13})=(3.07~a_0,\,4.97~a_0,\,4.97~a_0)$, where $a_0 \approx
52.91772105$~pm is the atomic unit of length, we observe a difference of only 0.3\% between the d4Z
and d6Z basis sets. According to the above discussion, such error is negligible in the present
context. A similar behavior is found for other tested geometries, see Table~\ref{tab:basis_conv}. As
a consequence, we recommend to use the results obtained with d4Z basis set for all 734
configurations. While calculations within the d5Z basis for $734$ configurations would be feasible
with our resources, we expect them to be at least an order of magnitude more costly than for d4Z. In
the face of minuscule accuracy improvement offered by the d5Z basis, as demonstrated in
Table~\ref{tab:basis_conv}, we refrained from its use for the whole surface.

To create an analytic representation of the surface of $|\mbm_3|^2$, we developed a model function expressed as a sum of a short-range part and a long-range part. The short-range part contains a factor which forces it to vanish for equilateral and centrosymmetric linear configurations. When the interatomic distances are sorted as described above, such configurations correspond to the situation where $r_{12}=r_{23}$ and $\theta_2$ is equal to either $\pi/3$ or $\pi$, respectively. 
The long-range part of the fitting function is a sum of squares of properly symmetrized expressions for the long-range expansion of the \textit{x}- and \textit{y}-components of the dipole moment
taken from the work of Li and Hunt.~\cite{Li97} The complete model function has the following form
\begin{widetext}
\begin{equation}\label{eq:dip}
\begin{split}
|\mbm_3|^2 
&= \left[|r_{12}-r_{23}| + (1-2\cos\theta_2)(1+\cos\theta_2)\right]\;
\sum_{l=1}^6 \e^{-\alpha_lr_{12}-\beta_lr_{23}} 
\sum_{\substack{0 \le i\le j\le k \le 5 \\ i+j+k\le7}}A_{ijk,l}\,r_{12}^ir_{23}^j(1+\cos k\theta_2)\\
&+\sum_{\alpha\in\{x,y\}}\left[S_{123}\left(
D_\mathrm{d}\frac{3{{R}_{12,\alpha}} \cos\theta_2+ {{R}_{23,\alpha}}}{r_{12}^3r_{23}^7}f_3(r_{12})f_7(r_{23})
+D_\mathrm{q}\frac{{{R}_{12,\alpha}}(5\cos^2\theta_2-1)+2 {{R}_{23,\alpha}}}{r_{12}^4r_{23}^6}f_4(r_{12})f_6(r_{23})\right)\right]^2,
\end{split}
\end{equation}
\end{widetext}
where $\alpha_l$, $\beta_l$, and $A_{ijk,l}$ are the fitting parameters,
$S_{123}$ denotes the operator that symmetrizes the expression it precedes 
with respect to the variables $\mathbf{r}_1$, $\mathbf{r}_2$, and $\mathbf{r}_3$, 
${R}_{ij,\alpha} = r_{i,\alpha} - r_{j,\alpha}$, 
$f_n(x)$ are the Tang--Toennies damping functions \cite{tang1984improved}
\begin{equation}\label{TTdamp}
f_n(x) = 1 - \e^{-x } \left(1 + x+\frac{x^2}{2!} +\dots +\frac{x^n}{n!}\right),
\end{equation}
and $D_\mathrm{d}$ and $D_\mathrm{q}$ are the asymptotic constants.

Several estimates of the the asymptotic constants $D_\mathrm{d}$ and $D_\mathrm{q}$  were reported in the literature.
Bruch \textit{et al.}~\cite{bruch1978dipole} provided $D_\mathrm{d}=-25.70$ and $D_\mathrm{q}=-5.18$
using pseudospectral representation of the dipole and quadrupole operators.
Somewhat different values were calculated by Li and Hunt,\cite{Li97}
$D_\mathrm{d}=-21.63$ and $D_\mathrm{q}=-4.36$, from the dipole and quadrupole integrals 
obtained by Fowler~\cite{fowler1990dispersion} through the coupled Hartree--Fock (CHF) method. 
Li and Hunt also provided alternative estimates of the $D_\mathrm{d}$
constant, $D_\mathrm{d}=-25.63$ and $D_\mathrm{d}=-25.4$,
using pseudospectral calculations of the dipole integrals
from the works of Bishop and Pipin~\cite{bishop1993calculation} 
and Whisnant and Byers Brown,~\cite{whisnant1973dispersion} respectively. However, the corresponding $D_\mathrm{q}$ constants were not given.
Martin\cite{Martin74} used a simple approximation to the resolvents based on the
calculations for hydrogen atoms and obtained $D_\mathrm{d}=-18.54$ and $D_\mathrm{q}=-3.70$. However, the last results are most likely not accurate enough for our purposes and are not considered in the following discussion.

\begin{table}
\caption{Comparison of the absolute relative percentage deviation,
$\big||\mbm_3|_\text{asympt}-|\mbm_3|_\textrm{calc}\big|/|\mbm_3|_\textrm{calc}\times100\%$, 
of the asymptotic expansion for the dipole moment
calculated using asymptotic constants 
from Ref.~\onlinecite{bruch1978dipole} ($D_\mathrm{d}=-25.70,\;D_\mathrm{q}=-5.18$) 
and from Ref.~\onlinecite{Li97} ($D_\mathrm{d}=-21.63,\;D_\mathrm{q}=-4.36$) 
with respect to the \textit{ab initio} results calculated as CC3[d4Z] and $^s$FCI[d4Z] (see text for definitions).
Interatomic distances $r_{ij}$ are in units of $a_0$.}
\label{tab:asympt}
\begin{ruledtabular}
\begin{tabular}{z{2.6}z{2.6}z{2.1}z{2.4}z{2.4}z{2.4}z{2.4}}
\multicolumn{3}{c}{}&
\multicolumn{2}{c}{CC3[d4Z]}&
\multicolumn{2}{c}{$^s$FCI[d4Z]}\\
\cline{4-5}
\cline{6-7}
\multicolumn{1}{c}{$r_{12}$} &
\multicolumn{1}{c}{$r_{23}$} &
\multicolumn{1}{c}{$r_{13}$} & 
\multicolumn{1}{c}{Ref.~\onlinecite{bruch1978dipole}} & 
\multicolumn{1}{c}{Ref.~\onlinecite{Li97}} & 
\multicolumn{1}{c}{Ref.~\onlinecite{bruch1978dipole}} &
\multicolumn{1}{c}{Ref.~\onlinecite{Li97}}\\
\hline
  8.012274&   8.012274&  13.5&   1.06\%&  14.94\%&    7.03\%&  21.75\%\\
  8.309025&   8.309025&  14.0&   1.39\%&  17.00\%&    9.08\%&  23.48\%\\
  8.605776&   8.605776&  14.5&   2.68\%&  18.09\%&   10.14\%&  24.37\%\\
  8.902527&   8.902527&  15.0&   3.22\%&  18.55\%&   10.57\%&  24.73\%\\
  9.199278&   9.199278&  15.5&   3.33\%&  18.64\%&   10.63\%&  24.79\%\\
  9.496029&   9.496029&  16.0&   3.14\%&  18.48\%&   10.31\%&  24.51\%\\
  9.792780&   9.792780&  16.5&   2.79\%&  18.19\%&   10.05\%&  24.29\%\\
 10.089531&  10.089531&  17.0&   2.35\%&  17.82\%&    9.68\%&  23.98\%\\
 10.386282&  10.386282&  17.5&   2.26\%&  17.73\%&    9.61\%&  23.93\%\\
 10.683033&  10.683033&  18.0&   1.38\%&  17.00\%&    8.82\%&  23.26\%\\
 10.979784&  10.979784&  18.5&   0.92\%&  16.61\%&    8.41\%&  22.92\%\\
 11.276534&  11.276534&  19.0&   0.47\%&  16.22\%&    8.02\%&  22.58\%\\
 11.573285&  11.573285&  19.5&   0.05\%&  15.88\%&    7.65\%&  22.28\%\\
\end{tabular}
\end{ruledtabular}
\end{table}

To determine which set of asymptotic constants is consistent with our \emph{ab initio} results 
for the dipole moment, in Table~\ref{tab:asympt} we compare our results obtained for isosceles geometries with large sides with the long-range expansion calculated using asymptotic constants from the papers of Bruch \textit{et al.}~\cite{bruch1978dipole} and Li and Hunt.\cite{Li97} However, one has to be careful in such direct comparison -- while our results are well-saturated with respect to the basis set size, higher-order excitations not included in the CC3 model are still missing. Therefore, there is a possibility of accidental agreement between two sets of results. To eliminate such possibility, we compare the long-range expansions not only with our CC3 data, but also with an estimate of the FCI results derived from our calculations. Similarly as at the CC3 level of theory, we expect the FCI results to be converged with the basis set size already in the d4Z basis. The FCI result in the d4Z basis is estimated by a simple scaling procedure employing results from the a3Z basis set\cite{a3_2022} as follows
\begin{equation}
^s\mathrm{FCI[d4Z]}=\mathrm{CC3[d4Z]}\times\frac{\mathrm{FCI[a3Z]}}{\mathrm{CC3[a3Z]}},
\end{equation}
where $M[B]$ denotes a result obtained using method $M$ with basis set $B$ 
and $^s$FCI[d4Z] is the approximation to the FCI result with the d4Z basis set. Raw \textit{ab initio} results CC3[d4Z] for all $734$ configurations considered in this work, and CC3[a3Z] and FCI[a3Z] results for configurations from Table~\ref{tab:asympt} are given in the Supplementary Material.

When the asymptotic constants from Bruch \textit{et al.}~\cite{bruch1978dipole} are used in the
asymptotic expansion, a good agreement with the CC3 dipole moments is observed, see
Table~\ref{tab:asympt}. The constants estimated by Li and Hunt\cite{Li97} lead to worse agreement
with CC3, with discrepancies of the order of 15\% or more. Crucially, similar trends are observed
when comparing with the FCI estimates. When the higher excitations are included, the differences
between the asymptotic expansion and \textit{ab initio} data rise to about 8--10\% in case of the
constants from Bruch \textit{et al.}~\cite{bruch1978dipole} The constants from Li and
Hunt~\cite{Li97} result in differences on the order of 22--25\%. Overall, for all isosceles
geometries with large sides (the smallest interatomic distance is 9.0~$a_0$), the mean absolute
relative deviation of the asymptotic expansions from the \emph{ab initio} data is $5$\% when the
asymptotic constants from Bruch \textit{et al.}~\cite{bruch1978dipole} are used and about $20$\%
while using the constants from Li and Hunt.\cite{Li97} For these reasons, the asymptotic constants
$D_\mathrm{d}$ and $D_\mathrm{q}$ from Bruch \textit{et al.}~\cite{bruch1978dipole} were used in the
model function, Eq.~(\ref{eq:dip}), in all subsequent calculations.

Similar analysis as in the previous paragraph can be used to estimate the uncertainty of our calculation
due to missing higher-order excitations. Based on the analysis presented in Table~\ref{tab:asympt},
and similar for other configurations, we assumed a very conservative estimation of the uncertainty
of our \emph{ab initio} data to be equal to 30\% of the recommended values. This estimation is
intended to take into account the basis set incompleteness error, the uncertainty of the value of
the asymptotic constants, and the missing contribution of the higher electronic excitations. We
believe that it is reasonable to interpret this estimated uncertainty as an expanded ($k = 2$)
uncertainty, roughly corresponding to a 95\% confidence interval.

After fixing the $D_\mathrm{d}$ and $D_\mathrm{q}$ constants, the remaining parameters in
Eq.~(\ref{eq:dip}) were fitted using the Levenberg--Marquardt algorithm as implemented in the SciPy
package.~\cite{2020SciPy-NMeth} During the optimization, the linear parameters (for fixed values of
the nonlinear ones) were obtained using the weighted linear least-squares minimization with weights
equal to inverse of estimated uncertainty squared.  The mean relative error of the fit is 23\% with
a median equal to 12\% with respect to the recommended values of the magnitude of the dipole moment
squared. In the Supplementary Material we provide the optimized parameters of the fitting
function~(\ref{eq:dip}) along with a FORTRAN code for its evaluation.

\section{{\em Ab initio} calculation of dielectric virial coefficients}

The theoretical framework enabling the {\em ab initio} calculation of the
dielectric virial coefficients appearing in the generalized
Clausius--Mossotti equation~(\ref{eq:Vir:eps}) has been developed in
Ref.~\onlinecite{Garberoglio_2021}. In this section, we will briefly recall
the most important results and describe the details of the calculations
reported below.

Denoting by $Q_N(V,T,E_0)$ the partition function of $N$ particles in a
volume $V$ at temperature $T$ and in the presence of an electric
field $E_0$ generated by sources external to the volume $V$, and defining
the reduced partition functions
\begin{equation}
  \frac{Z_N(T,E_0)}{N!} = \frac{Q_N(V,T,E_0) ~ V^N}{Q_1(V,T,E_0)^N},
  \label{eq:ZN}
\end{equation}
the dielectric virial coefficients appearing in Eq.~(\ref{eq:Vir:eps}) are given by
\begin{eqnarray}
  B_\varepsilon &=& \frac{2 \pi \kB T}{3 V}
  \frac{\partial^2 Z_2(V,T,E_0)}{\partial E_0^2},
  \label{eq:Beps} \\
  C_\varepsilon &=& -\frac{2 \pi \kB T}{3} \left[
    \frac{2}{V^2} \left( \frac{\partial Z_2}{\partial E_0} \right)^2 +
    \frac{2(Z_2-V^2)}{V^2} \frac{\partial^2 Z_2}{\partial E_0^2}
    \right. \nonumber \\
  & & \left. -\frac{1}{3V} \left(
  \frac{\partial^2 Z_3}{\partial E_0^2} - 3 V \frac{\partial^2
    Z_2}{\partial E_0^2}
  \right) \right],
  \label{eq:Ceps}  
\end{eqnarray}
where the derivatives with respect to the external electric field are to be
evaluated in the limit $E_0 \to 0$. In this work, we will be concerned with
the third dielectric virial coefficient, $C_\varepsilon(T)$ of
Eq.~(\ref{eq:Ceps}). In the general case of quantum
statistical mechanics, one can write~\cite{Ceperley1995}
\begin{eqnarray}
  Q_N(V,T,E_0) = \frac{1}{N!} \sum_{i,\sigma} \langle i | \mathrm{e}^{-\beta H(N)}
  {\mathcal P}_\sigma |i \rangle,
  \label{eq:QN}
\end{eqnarray}
where $\beta = 1/(\kB T)$, $\sigma$ runs over the permutations of $N$
objects, and ${\mathcal P}_\sigma$ is 
the corresponding operator in the Hilbert space of $N$ particles, which
includes the sign of the permutation in the case of fermions. The sum
over $i$ in Eq.~(\ref{eq:QN}) runs over all the eigenstates of the
Hamiltonian $H(N)$, which can be in turn expressed as $H(N) = H_0(N) +
\Delta H_\mbm(N) + \Delta H_{\mbalpha}(N)$ where
\begin{eqnarray}
  H_0(N) &=& \sum_{i=1}^N \frac{\mbpi_i^2}{2m} +
  \sum_{i<j} u_2(i,j) + \sum_{i<j<k} u_3(i,j,k) + \ldots,
  \label{eq:H0} \\
  \Delta H_\mbm(N) &=& - \left(
  \sum_{i=1}^N \mbm_1(i) + \sum_{i<j} \mbm_2(i,j) + \right . \nonumber \\
  & & \left.
  \sum_{i<j<k} \mbm_3(i,j,k) + \ldots  \right) \cdot \mbE_0 \label{eq:DH1} \\
  \Delta H_{\mbalpha}(N) &=& -\frac{1}{2} \mbE_0 \cdot \left(
  \sum_{i=1}^N \mbalpha_1(i) + \sum_{i<j} \mbalpha_2(i,j) + \right.
  \nonumber \\
  & & \left. 
  \sum_{i<j<k} \mbalpha_3(i,j,k) + \ldots \right) \cdot \mbE_0.
  \label{eq:DH2}
\end{eqnarray}
In Eqs.~(\ref{eq:H0})--(\ref{eq:DH2}), $m$ denotes the mass of the atoms, $\mbpi$ is the
momentum operator, and $u_k(1,\ldots,k)$, $\mbm_k(1,\ldots,k)$ and
$\mbalpha_k(1,\ldots,k)$ are the $k$-body potential, dipole moment, and
polarizability, respectively. Notice that the polarizabilities are $3
\times 3$ matrices.

In this work, we used for $u_2$ the pair potential developed by
Czachorowski {\em et al.},~\cite{u2_2020} while $u_3$ is given by the three-body potential
developed by Lang {\em et al.}~\cite{u3_2023} We further used the pair
polarizability $\mbalpha_2$ by Cencek {\em et al.}~\cite{Cencek11} and the
three-body polarizability by Lang {\em et al.}~\cite{a3_2022} All of these
quantities come with rigorous uncertainty estimates.
In the case of noble gases, the first non-zero dipole-moment contribution
in Eq.~(\ref{eq:DH1}) comes from the term $\mbm_3$, which was determined
for this work and is reported in Sec.~\ref{sec:dipole}. 
Effects beyond the nonrelativistic Born--Oppenheimer approximation result in the interaction among
${}^3$He atoms being slightly different than that among ${}^4$He atoms. This has been taken into
account in the case of the pair potential,~\cite{u2_2020} but neglected for other quantities since
the effect of the isotopic difference is negligible compared to other uncertainties.

For fluids like helium, where quantum effects are significant, the evaluation of $C_\varepsilon(T)$
from Eq.~(\ref{eq:Ceps}), using Eqs.~(\ref{eq:ZN}) and (\ref{eq:QN}), is most conveniently performed
using the path-integral formulation of quantum statistical
mechanics.~\cite{FH,Garberoglio2011a,Garberoglio_2021} In this approach, one can show that the
reduced partition functions $Z_N$ can be evaluated as classical partition functions of an equivalent
system, where each of the $N$ atoms is substituted by a ring polymer with $P$ monomers. The
equivalence is exact in the $P \to \infty$ limit, although convergence of the results is usually
obtained with a finite value of $P$, which is generally larger the lower the temperature
considered. Additionally, the formalism provides a well-defined interaction law between the ring
polymers corresponding to the various atoms, which depends on the actual interaction potentials
$u_k(1,\ldots,k)$; successive monomers interacts with a harmonic potential, with a spring constant
that depends on $T$ and $P$, as well as on the atomic mass and temperature.  Physically, the size of
the polymers is proportional to the thermal de~Broglie wavelength of the atoms, $\Lambda = h /
\sqrt{2\pi m \kB T}$, and takes into account quantum diffraction effects due to the uncertainty
principle.
Classical many-particles configurations obtained using the coordinates of monomers with the same
index $p$ ($1 \leq p \leq P$) are usually called an ``imaginary-time slice''.
Exchange effects due to particle indistinguishability are taken into account by
considering coalesced ring polymers according to the specific permutation appearing in
Eq.~(\ref{eq:QN}). As discussed in Refs.~\onlinecite{Garberoglio_2021} and
\onlinecite{Garberoglio2011a}, the $6$ possible permutations of $3$ particles appearing in the sum
of $\sigma$ in Eq.~(\ref{eq:QN}) can be divided into three classes. The first class, with only one
element, is the identity permutation, which corresponds to Boltzmann statistics (that is,
distinguishable particles), and that will be denoted by the symbol $\therefore$. The second class
includes the three permutations where only one pair is exchanged, and will be denoted by $\cdot
|$. Finally, the third class includes the two cyclic permutations and we will denote it by
$\triangle$.  In the case of two particles, one has only the identity ($:$) and the pair permutation
($|$). As a consequence, one can write
\begin{eqnarray}
  Z_3 &=& Z_3^\therefore + Z_3^{\cdot |} + Z_3^\triangle \\
  Z_2 &=& Z_2^: + Z_2^|,
\end{eqnarray}
and, using Eq.~(\ref{eq:Ceps}), derive three contributions to the third
dielectric virial coefficient~\cite{Garberoglio_2021}
\begin{widetext}
  \begin{small}
  \begin{eqnarray}
    C_\varepsilon^\therefore(T) &=& \frac{2\pi}{3} \int 
    \left[ \frac{1}{3}\left\langle
      \left(\frac{\beta |\overline{\mbm_3^\therefore}|^2}{3} +
      \overline{A_3^\therefore} \right) \mathrm{e}^{-\beta
        \overline{V_3^\therefore}} - \sum_{i<j}
      \overline{\alpha_\mathrm{iso}^:}(\mbr_{ij}) \mathrm{e}^{-\beta
        \overline{V_2^:}(\mbr_{ij})} \right\rangle - 2 \langle \mathrm{e}^{-\beta
        \overline{V_2^:}(\mbr_{21})}-1 \rangle \langle \overline{\alpha_\mathrm{iso}^:}(\mbr_{31}) \mathrm{e}^{-\beta \overline{V_2^:}(\mbr_{31})}
      \rangle \right] ~ \mathrm{d}\mbr_2 \mathrm{d}\mbr_3,
    \label{eq:CepsB} \\
    C^{\cdot |}_\varepsilon(T) &=& \frac{(-1)^{2I}}{2I+1} \frac{2\pi}{3}
    \frac{\Lambda^3}{2^{3/2}}
    \int \mathrm{d}\mbr ~ \left[
    \left\langle \left(
    \frac{\beta \left|\overline{\mbm_3^{\cdot |}}\right|^2}{3} + 
    \overline{A_3^{\cdot |}} \right)
    \mathrm{e}^{-\beta \overline{V_3^{\cdot |}}}
    - \overline{\alpha_\mathrm{iso}^|} \mathrm{e}^{-\beta \overline{V_2^|}}
    \right\rangle  - 
    2 \langle \mathrm{e}^{-\beta \overline{V_2^|}} \rangle
    \langle \overline{\alpha_\mathrm{iso}^:} \mathrm{e}^{-\beta
      \overline{V_2^:}} \rangle 
    -
    2 \langle \overline{\alpha_\mathrm{iso}^|} \mathrm{e}^{-\beta \overline{V_2^|}} \rangle
    \langle \mathrm{e}^{-\beta \overline{V_2^:}} -1 \rangle \right]
    \label{eq:Ceps1p} \\
    C^\triangle_\varepsilon(T) &=& \frac{2 \pi}{3}
    \frac{\Lambda^6}{(2I+1)^2} \left[
    \frac{2}{3^{5/2}} \left\langle
    \left(
    \frac{\beta \left| \overline{\mbm_3^\triangle} \right|^2}{3} + 
    \overline{A_3^\triangle} \right)
    \mathrm{e}^{-\beta \overline{V_3^\triangle}}
    \right\rangle
    -
    \frac{1}{4} \left\langle \mathrm{e}^{-\beta \overline{V_2^|}} \right\rangle
    \left\langle \overline{\alpha_\mathrm{iso}^|} \mathrm{e}^{-\beta
      \overline{V_2^|}} \right\rangle 
    \right],  \label{eq:CepsCyc}  
  \end{eqnarray}    
  \end{small}
\end{widetext}
where we have denoted by $I$ the nuclear spin of the atoms under consideration, that is $I=0$ for
${}^4$He and $I=1/2$ for ${}^3$He.  The horizontal bars denote the average values of the various
observables on the configurations in each imaginary-time slice of the ring polymers, whereas angular
brackets denote the average over the possible internal configuration of polymers. In
Eq.~(\ref{eq:CepsB}), the first monomer of one polymer is fixed at the origin of the coordinate
system $\mbr_1 = 0$, the vectors $\mbr_2$ and $\mbr_3$ denote the position of the first monomer of
the two polymers corresponding to the other two quantum particles and we have defined $\mbr_{ij} =
\mbr_i - \mbr_j$. In Eq.~(\ref{eq:Ceps1p}), the vector $\mbr$ points from the first
monomer of the coalesced polymer to the first monomer of the particle unaffected by the permutation.
The averages in Eq.~(\ref{eq:CepsCyc}) are obtained over the configuration of fully coalesced
polymers of either three ($\triangle$) or two ($|$) particles.

The other quantities appearing in Eqs.~(\ref{eq:CepsB})--(\ref{eq:CepsCyc})
are: the isotropic component of the pair polarizability,
$\alpha_\mathrm{iso} = \mathrm{tr}[\mbalpha_2]/3$, the isotropic component
of the full three-body polarizability
\begin{eqnarray}
  A_3(\mbr_{12},\mbr_{13},\mbr_{23}) &=& \mathrm{tr}\left[ \mbalpha_3(\mbr_{12},\mbr_{13},\mbr_{23}) +
    \mbalpha_2(\mbr_{12},\mbr_{13}) + \right. \nonumber \\
    & & \left. 
  \mbalpha_2(\mbr_{12},\mbr_{23}) + \mbalpha_2(\mbr_{13},\mbr_{23}) \right]/3,  
\end{eqnarray}
the pair potential ($V_2 = u_2$) and the three-body
potential $V_3(\mbr_{12},\mbr_{13},\mbr_{23}) = u_3(\mbr_{12},\mbr_{13},\mbr_{23}) + u_2(\mbr_{12},\mbr_{13}) +
u_2(\mbr_{12},\mbr_{23}) + u_2(\mbr_{13},\mbr_{23})$.

For details on how to sample the ring-polymer configurations corresponding to the various
permutations, we refer to our original
publications~\cite{Garberoglio2009b,Garberoglio2011,Garberoglio2011a} as well as the seminal papers
by Fosdick and Jordan.~\cite{Fosdick1966,Jordan1968} In performing the calculations solving
Eqs.~(\ref{eq:CepsB})--(\ref{eq:CepsCyc}), we used a temperature-dependent value of $P$, according
to $P = \mathrm{nint}(7 + 1600~\mathrm{K}/T)$ for ${}^4$He and $P = \mathrm{nint}(7 +
2000~\mathrm{K}/T)$ for ${}^3$He, where $\mathrm{nint}(x)$ denotes the nearest integer to $x$. For
the convenience of calculation, we split the expressions for the various components of
$C_\varepsilon(T)$ into five contributions, (roughly) in order of decreasing magnitude: the first,
denoted by $C_\varepsilon^\mathrm{\therefore, 2-body}$, is the Boltzmann contribution depending only
on pair properties (that is, $u_2$ and $\mbalpha_2$). The second term, denoted by,
$C_\varepsilon^\mathrm{\therefore, diff-32}$, is the Boltzmann contribution that takes into account
the difference between the dielectric virial coefficient calculated with the non-additive three-body
potential and polarizability and $C_\varepsilon^\mathrm{\therefore,2-body}$. The two following
contributions are the odd and even exchange contributions -- $C_\varepsilon^{\cdot |}$ and
$C_\varepsilon^{\triangle}$, respectively -- from Eqs.~(\ref{eq:Ceps1p}) and
(\ref{eq:CepsCyc}). These four terms are computed neglecting the contribution of the three-body
dipole-moment surface. The remainder, which we denote by $C_\varepsilon^{\mathrm{dip}}$, takes into
account all the terms where the dipole-moment surface $\mbm_3$ appears explicitly in both the
Boltzmann and exchange contributions.  This approach was found to be convenient in the calculation
of density virial coefficients because the difference terms, which depend on the computationally
demanding three-body surfaces, can be evaluated with an accuracy comparable to that of the two-body
term by sampling a smaller number of Monte Carlo configurations.  Notice that
$C_\varepsilon^{\mathrm{dip}}$ depends on the average of the directions of the dipole moment across
the ring-polymer beads. However, the dipole-moment surface developed here, Eq.~(\ref{eq:dip}),
provides only the squared modulus of the dipole moment. We approximated the direction of the dipole
moment by that provided by the long-range contribution developed by Li and Hunt~\cite{Li97} using
the parameters $D_\mathrm{d}$ and $D_\mathrm{q}$ computed by Bruch {\em et
  al.}~\cite{bruch1978dipole} We performed as many independent runs of $10^6$ Monte Carlo samples as
required to obtain a statistical uncertainty smaller than 60\% of the propagated uncertainty from
the potential and polarizability, which has been evaluated according to the method discussed in
Sec.~\ref{sec:unc}. In order to reach this target, we needed at least 30 independent runs at each
temperature.

\section{Propagation of the uncertainty}
\label{sec:unc}

Apart from the statistical uncertainty of the Monte Carlo calculation,
which can be in principle reduced at will given a sufficient supply of
computational power, there are five other sources of uncertainty for
$C_\varepsilon(T)$, which come from the propagation of the uncertainty in
the potential-energy surfaces $u_2$ and $u_3$, the polarizabilities
$\mbalpha_2$ and $\mbalpha_3$, and the dipole moment $\mbm_3$.
The standard way to estimate these uncertainties is to perform calculations
of $C_\varepsilon(T)$ using perturbed
quantities,~\cite{Garberoglio2009b,Garberoglio2011} ({\em e.g.}, $u_3 + \delta u_3$ 
and $u_3 - \delta u_3$), obtaining perturbed values of the virial
coefficients, say $C_\varepsilon^+$ and $C_\varepsilon^-$. Assuming that the
uncertainty $\delta u_3$ is provided at coverage factor $k=2$, the estimate
of the standard uncertainty of the dielectric virial coefficient coming from the
pair potential would be given by
\begin{equation}
  \delta C_\varepsilon^{[u_3]} = \frac{1}{4}
  \left| C_\varepsilon^+ - C_\varepsilon^- \right|,
  \label{eq:dC_old}
\end{equation}
with analogous expressions for the two-body potential, the
polarizabilities, and the dipole moment.

The main drawback of this approach is that the value of the propagated
uncertainty $\delta C_\varepsilon$ is not known in
advance, and this requires performing long calculations of the perturbed
virials so that the difference in Eq.~(\ref{eq:dC_old}) becomes smaller
than the combined statistical uncertainties of $C_\varepsilon^+$ and
$C_\varepsilon^-$. 
In a recent paper,~\cite{Garberoglio2021a} we put forward an alternative
way to estimate $\delta C_\varepsilon(T)$, based on functional
differentiation of the expressions leading to the virial coefficient with
respect to potential, polarizability, and dipole-moment surfaces. We used
the classical expressions for the virial coefficients, and we evaluated the resulting
expressions using the fourth-order Feynman--Hibbs correction to the pair
potential in order to take into account quantum effects. 

In this paper, we extend this approach by taking directly the functional
derivatives of the path-integral expressions, that is
Eqs.~(\ref{eq:CepsB})--(\ref{eq:CepsCyc}). In this case, quantum effects
are considered from the very start and we can safely proceed to evaluate
propagated uncertainties down to very low temperatures, where a
semiclassical approach would most likely fail. Additionally, we can include
the uncertainty of exchange terms in a rigorous way.  Since the third
dielectric virial coefficient is given by a sum of three terms, each of
which depends on five different surfaces, there are 15
contributions to the overall uncertainty. We will discuss here only a few
of them to point out the main features of the calculations.  All of the
contributions to the overall uncertainty are then summed in quadrature to
produce the overall value of the propagated uncertainty.
Explicit expressions for all 15 contributions to the uncertainty of
$C_\varepsilon$ are given in the Supplementary Material.

In calculating the variation of the third dielectric virial coefficient with respect to a quantity
$s$ with uncertainty $\delta s$ ($s = u_2, u_3, \alpha_2, \alpha_3, \mbm_3$), which depends on some
variables $\mbX$ ($\mbX$ is the radial distance in the case of $u_2$ and $\alpha_2$, or the three
interatomic distances in the case of $u_3, \alpha_3$, and $\mbm_3$), one obtains for the propagated
standard uncertainty the general formula
\begin{equation}
  \delta C_\varepsilon^{[s]} = \frac{1}{2} \int \delta s \left| \frac{\delta
    C_\varepsilon}{\delta s} \right| \diff\mbX,
  \label{eq:unc_prop}
\end{equation}
where the factor of $2$ takes into account the fact that $\delta s$ is an
expanded uncertainty with coverage factor $k=2$.
Since the functional derivative $\delta C_\varepsilon/\delta s$ can in
general have positive and negative values, we adopted the
conservative choice of taking the absolute value of the integrand in Eq.~(\ref{eq:unc_prop}). 

Let us consider first how to propagate the uncertainty due to the
isotropic part of the three-body polarizability $\alpha_3 =
\mathrm{tr}[\mbalpha_3]/3$ to the value 
of the Boltzmann component of the third dielectric virial
coefficient. Inserting 
Eq.~(\ref{eq:CepsB}) in Eq.~(\ref{eq:unc_prop}), one obtains, for the
standard uncertainty, 
\begin{equation}
  \delta C_\varepsilon^{\therefore [\alpha_3]}  = \frac{1}{2}
  \frac{2\pi}{9} \int \left | \left\langle
  \overline{\delta \alpha_3^\therefore} \e^{-\beta \overline{V_3^\therefore}}
  \right\rangle \right| \diff\mbr_2 \diff\mbr_3,
  \label{eq:dCepsa3}
\end{equation}
and similarly, in the case of $u_3$,
\begin{equation}
  \delta C_\varepsilon^{\therefore [u_3]}  = \frac{1}{2}
  \frac{2\pi \beta}{9} \int \left | \left\langle
  \overline{\delta u_3^\therefore} ~ 
  \overline{A_3^\therefore} \e^{-\beta \overline{V_3^\therefore}}
  \right\rangle \right| \diff\mbr_2 \diff\mbr_3.
  \label{eq:dCepsu3}  
\end{equation}

By performing the functional derivative of Eq.~(\ref{eq:CepsB}) with respect to the
isotropic component of the pair polarizability and the pair potential, we
obtain an expression for their contribution to the overall uncertainty
\begin{widetext}
  \begin{eqnarray}
  \delta C_\varepsilon^{\therefore [\alpha_2]}  &=& \frac{1}{2}
  \frac{2\pi}{9} \int \left | \left\langle
  \sum_{i<j} \overline{\delta \alpha_\mathrm{iso}}(\mbr_{ij}) \left(
  \e^{-\beta \overline{V_3^\therefore}} - \e^{-\beta \overline{V_2^\therefore}(\mbr_{ij})}
  \right)
  - 6 \left( \e^{-\beta \overline{V_2^\therefore}(\mbr_{12})} - 1 \right)
  \overline{\delta \alpha_\mathrm{iso}}(\mbr_{13}) \e^{-\beta \overline{V_2^\therefore}(\mbr_{13})}
  \right\rangle \right| \diff\mbr_2 \diff\mbr_3
  \label{eq:dCepsa2} \\
  \delta C_\varepsilon^{\therefore [u_2]}  &=& \frac{1}{2}
  \frac{2\pi \beta}{9} \int \left | \left\langle
  \sum_{i<j} \overline{\delta u_2}(\mbr_{ij}) \left[
    \left(\frac{\beta |\overline{\mbm_3^\therefore}|^2}{3} +
    \overline{A_3^\therefore} \right) \mathrm{e}^{-\beta
      \overline{V_3^\therefore}}
    - \overline{\alpha_\mathrm{iso}}(\mbr_{ij}) \e^{-\beta \overline{V_2^\therefore}(\mbr_{ij})}
    \right] -
  \right. \right. \nonumber \\
  & &
  \left. \left.
  6 \overline{\delta u_2}(\mbr_{12}) \e^{-\beta \overline{V_2^\therefore}(\mbr_{12})}
  \left(
  \e^{-\beta \overline{V_2^\therefore}(\mbr_{13})}
  (\overline{\alpha_\mathrm{iso}}(\mbr_{12}) +
  \overline{\alpha_\mathrm{iso}}(\mbr_{13}))
  -  \overline{\alpha_\mathrm{iso}}(\mbr_{12})
  \right)
  \right\rangle \right| \diff\mbr_2 \diff\mbr_3,
  \label{eq:dCepsu2}  
  \end{eqnarray}
\end{widetext}
and, finally, functional differentiation with respect to the dipole moment
gives
\begin{equation}
  \delta C_\varepsilon^{\therefore [\mbm_3]}  = \frac{1}{2}
  \frac{4\pi \beta}{27} \int \left | \left\langle
  \overline{\delta\mbm_3^\therefore} \cdot \overline{\mbm_3^\therefore}
  \e^{-\beta \overline{V_3^\therefore}}  
  \right\rangle \right| \diff\mbr_2 \diff\mbr_3.
  \label{eq:dCepsm}  
\end{equation}

We notice that the last terms in Eqs.~(\ref{eq:dCepsa2}) and
(\ref{eq:dCepsu2}) explicitly involve the pairs 1--2 and 1--3. The
integral, however, does not depend on the labels assigned to the particles
and there are six possible ways to extract from three particles two pairs
that have one label in common. In actual calculations, it is convenient to
average over the symmetrization in all the possible pairs, which
will conveniently remove the factor of $6$ in front of the original
expression.  Furthermore, the values of the propagated uncertainties
converge much faster than the dielectric virial calculations. We found that
it was sufficient to use $P = \mathrm{nint}(4 + 140~\mathrm{K}/T)$ with 200~000 Monte
Carlo samples for both isotopes in order to have well-converged results.

\section{Results and discussion}

\begin{table*}
  \caption{Third dielectric virial coefficient $C_\varepsilon$ for
    ${}^4$He, its expanded uncertainty $U(C_\varepsilon)$, together with the various
    contributions from the pair potential
    ($C_\varepsilon^\mathrm{\therefore,2-body}$), the three-body potential and
    polarizability ($C_\varepsilon^{\mathrm{\therefore,diff-32}}$), 
    the dipole moment  ($C_\varepsilon^\mathrm{dip}$), and the exchange terms of
    Eqs.~(\ref{eq:Ceps1p}) and (\ref{eq:CepsCyc}). $U(C_\varepsilon)$ are
    expanded uncertainties at $k=2$ and include the statistical uncertainty
    of the calculation as well as the propagated uncertainties from
    potentials, polarizabilities, and dipole moment.}
  \begin{tabular}{d|d|d||d|d|d|d|d}
    \multicolumn{1}{c|}{Temperature} &
    \multicolumn{1}{c|}{$C_\varepsilon$} &
    \multicolumn{1}{c||}{$U(C_\varepsilon)$} &
    \multicolumn{1}{c|}{$C_\varepsilon^{\mathrm{\therefore,2-body}}$} &
    \multicolumn{1}{c|}{$C_\varepsilon^{\mathrm{\therefore,diff-32}}$} &    
    \multicolumn{1}{c|}{$C_\varepsilon^\mathrm{dip}$} &
    \multicolumn{1}{c|}{$C^{\cdot |}_\varepsilon$} &
    \multicolumn{1}{c}{$C^\triangle_\varepsilon$} \\    
    \multicolumn{1}{c|}{(K)} &
    \multicolumn{1}{c|}{(cm${}^9$~mol${}^{-3}$)} & 
    \multicolumn{1}{c||}{(cm${}^9$/mol${}^3$)} &
    \multicolumn{1}{c|}{(cm${}^9$/mol${}^3$)} &
    \multicolumn{1}{c|}{(cm${}^9$/mol${}^3$)} &
    \multicolumn{1}{c|}{($10^{-5}$~cm${}^9$/mol${}^3$)} &    
    \multicolumn{1}{c|}{(cm${}^9$/mol${}^3$)} &   
    \multicolumn{1}{c}{(cm${}^9$/mol${}^3$)} \\
    \hline
1	&	-3.89	&	0.55	&	-1.832	&	-0.628	&	531.0	&	-0.5518	&	-0.8852	\\
1.2	&	-2.26	&	0.35	&	-1.165	&	-0.465	&	433.6	&	-0.1541	&	-0.4840	\\
1.4	&	-1.44	&	0.23	&	-0.744	&	-0.399	&	312.0	&	-0.0435	&	-0.2556	\\
1.6	&	-1.07	&	0.18	&	-0.572	&	-0.348	&	250.4	&	-0.0153	&	-0.1392	\\
1.8	&	-0.83	&	0.14	&	-0.448	&	-0.322	&	205.8	&	0.0128	&	-0.0786	\\
2	&	-0.72	&	0.11	&	-0.388	&	-0.292	&	169.2	&	0.0090	&	-0.0485	\\
2.1768	&	-0.62	&	0.10	&	-0.319	&	-0.288	&	146.3	&	0.0163	&	-0.0304	\\
2.5	&	-0.51	&	0.08	&	-0.236	&	-0.268	&	114.7	&	0.0111	&	-0.0136	\\
3	&	-0.39	&	0.06	&	-0.173	&	-0.224	&	85.6	&	0.0053	&	-0.0042	\\
3.5	&	-0.34	&	0.05	&	-0.129	&	-0.217	&	68.1	&	0.0025	&	-0.0014	\\
4	&	-0.31	&	0.05	&	-0.111	&	-0.198	&	56.3	&	0.0017	&	-0.0005	\\
4.222	&	-0.29	&	0.04	&	-0.101	&	-0.193	&	51.7	&	0.0009	&	-0.0003	\\
4.5	&	-0.28	&	0.04	&	-0.095	&	-0.188	&	47.2	&	0.0007	&	-0.0002	\\
5	&	-0.27	&	0.04	&	-0.083	&	-0.186	&	40.8	&		&		\\
6	&	-0.24	&	0.03	&	-0.075	&	-0.168	&	32.3	&		&		\\
7	&	-0.23	&	0.03	&	-0.068	&	-0.166	&	26.7	&		&		\\
8	&	-0.22	&	0.03	&	-0.066	&	-0.157	&	22.9	&		&		\\
10	&	-0.23	&	0.02	&	-0.070	&	-0.156	&	18.1	&		&		\\
12	&	-0.22	&	0.02	&	-0.074	&	-0.148	&	15.3	&		&		\\
13.8031	&	-0.23	&	0.02	&	-0.080	&	-0.148	&	13.5	&		&		\\
15	&	-0.23	&	0.02	&	-0.083	&	-0.146	&	12.6	&		&		\\
17	&	-0.23	&	0.02	&	-0.090	&	-0.145	&	11.4	&		&		\\
20	&	-0.24	&	0.02	&	-0.100	&	-0.140	&	10.3	&		&		\\
24.5561	&	-0.25	&	0.02	&	-0.114	&	-0.134	&	9.1	&		&		\\
30	&	-0.27	&	0.02	&	-0.132	&	-0.134	&	8.3	&		&		\\
40	&	-0.29	&	0.02	&	-0.159	&	-0.132	&	7.5	&		&		\\
50	&	-0.32	&	0.02	&	-0.184	&	-0.134	&	7.2	&		&		\\
54.3584	&	-0.32	&	0.02	&	-0.193	&	-0.131	&	7.1	&		&		\\
60	&	-0.34	&	0.02	&	-0.205	&	-0.133	&	7.0	&		&		\\
75	&	-0.36	&	0.02	&	-0.233	&	-0.124	&	7.0	&		&		\\
83.8058	&	-0.37	&	0.02	&	-0.248	&	-0.125	&	7.1	&		&		\\
100	&	-0.39	&	0.02	&	-0.272	&	-0.123	&	7.2	&		&		\\
125	&	-0.43	&	0.02	&	-0.304	&	-0.124	&	7.6	&		&		\\
150	&	-0.45	&	0.02	&	-0.331	&	-0.119	&	8.0	&		&		\\
161.4	&	-0.46	&	0.02	&	-0.342	&	-0.119	&	8.2	&		&		\\
175	&	-0.47	&	0.02	&	-0.354	&	-0.114	&	8.4	&		&		\\
200	&	-0.49	&	0.03	&	-0.375	&	-0.114	&	8.8	&		&		\\
225	&	-0.50	&	0.03	&	-0.393	&	-0.111	&	9.2	&		&		\\
234.3156	&	-0.51	&	0.03	&	-0.399	&	-0.107	&	9.4	&		&		\\
250	&	-0.52	&	0.03	&	-0.409	&	-0.106	&	9.7	&		&		\\
273.16	&	-0.53	&	0.03	&	-0.423	&	-0.108	&	10.1	&		&		\\
300	&	-0.54	&	0.03	&	-0.438	&	-0.104	&	10.5	&		&		\\
302.9146	&	-0.54	&	0.03	&	-0.439	&	-0.103	&	10.6	&		&		\\
325	&	-0.55	&	0.03	&	-0.450	&	-0.099	&	10.9	&		&		\\
350	&	-0.56	&	0.03	&	-0.461	&	-0.098	&	11.4	&		&		\\
375	&	-0.57	&	0.03	&	-0.472	&	-0.098	&	11.8	&		&		\\
400	&	-0.58	&	0.03	&	-0.482	&	-0.098	&	12.2	&		&		\\
429.7485	&	-0.59	&	0.03	&	-0.492	&	-0.094	&	12.6	&		&		\\
450	&	-0.59	&	0.03	&	-0.499	&	-0.092	&	13.0	&		&		\\
500	&	-0.60	&	0.03	&	-0.515	&	-0.083	&	13.8	&		&		\\
600	&	-0.62	&	0.03	&	-0.541	&	-0.081	&	15.3	&		&		\\
700	&	-0.64	&	0.03	&	-0.562	&	-0.077	&	16.8	&		&		\\
800	&	-0.64	&	0.04	&	-0.579	&	-0.065	&	18.2	&		&		\\
900	&	-0.65	&	0.04	&	-0.593	&	-0.054	&	19.6	&		&		\\
1000	&	-0.66	&	0.04	&	-0.605	&	-0.055	&	21.0	&		&		\\
1500	&	-0.67	&	0.04	&	-0.644	&	-0.027	&	27.6	&		&		\\
2000	&	-0.66	&	0.05	&	-0.662	&	0.002	&	33.8	&		&		\\
2500	&	-0.65	&	0.05	&	-0.670	&	0.021	&	39.7	&		&		\\
3000	&	-0.62	&	0.06	&	-0.672	&	0.053	&	45.5	&		&		
  \end{tabular}      
  \label{tab:He4}
\end{table*}

The results of our calculations are reported in Tables~\ref{tab:He4} and
\ref{tab:He3} for the ${}^4$He and ${}^3$He isotopes, respectively.  The
two sets of data are mutually compatible, within expanded uncertainties,
down to $T \sim 4$~K. Below this temperature, quantum exchange effects contribute
significantly to the difference in $C_\varepsilon$ between the two
isotopes.

Table~\ref{tab:He4} also reports the values of all the contributions to
$C_\varepsilon$ that have been computed for this work: the value assuming
only pair interactions ($C_\varepsilon^\mathrm{\therefore,2-body}$), the contribution
due to three-body properties ($C_\varepsilon^\mathrm{\therefore,diff-32}$, which takes
into account the three-body potential and polarizability), the
dipole-moment contribution ($C_\varepsilon^\mathrm{dip}$, coming from the
terms proportional to $\mbm_3$ in
Eqs.~(\ref{eq:CepsB})--(\ref{eq:CepsCyc})), and the quantum statistical
(exchange) contributions at low temperature, Eqs.~(\ref{eq:Ceps1p}) and
(\ref{eq:CepsCyc}).  In general, the dipole-moment contribution is
completely negligible at all the temperatures investigated here, as already
noticed in our previous work~\cite{Garberoglio_2021} where we used an
approximate form for the dipole moment of the helium trimer due to Li and
Hunt.~\cite{Li97} We note, however, that the values for the dipolar
contribution that we obtain with the new dipole-moment surface are 3 to 20
times larger than the ones obtained with the surface of
Ref.~\onlinecite{Li97}.

\begin{figure}
\includegraphics[width=0.95\linewidth]{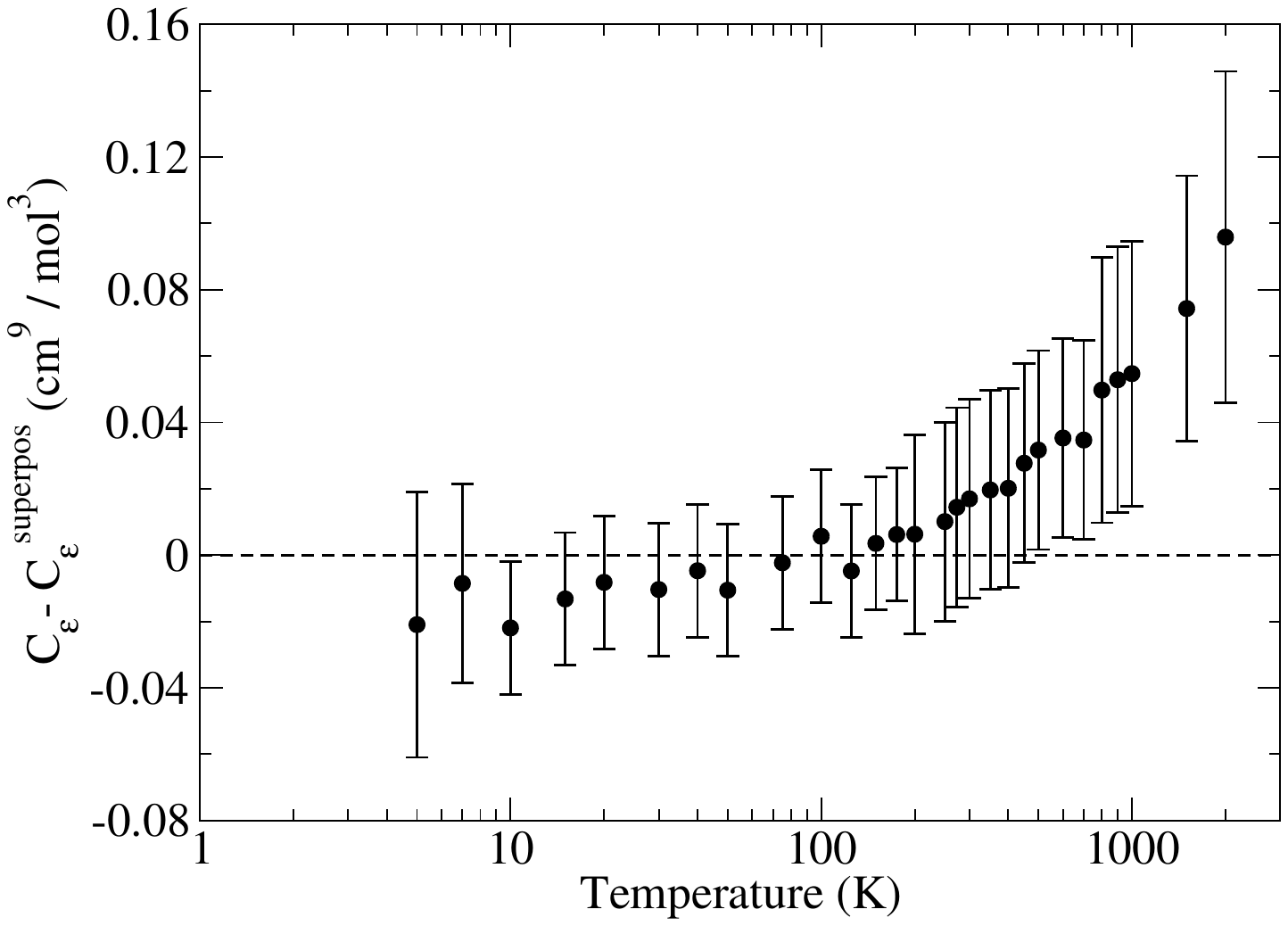}    
  \caption{The values of $C_\varepsilon$ for ${}^4$He computed in the present work using as a
    baseline the values computed using the superposition approximation for the three-body
    polarizability.~\cite{Garberoglio_2021} Error bars correspond to expanded $(k=2)$
    uncertainties.}
\label{fig:superpos}
\end{figure}

Quantum exchange contributions are found to be sizable (that is, comparable
to the overall uncertainty of the calculation) below $T \sim 3$~K for both
isotopes.
When compared with our previous calculations,~\cite{Garberoglio_2021} which used an approximate form
of the three-body polarizability, the present values of $C_\varepsilon$ are in reasonable agreement
for most of the temperature range. As shown in Fig.~\ref{fig:superpos}, there is systematic upwards
shift for $T>300$~K, and the difference becomes larger than the uncertainty of the present
calculation for $T>1000$~K. Given the minor contribution from the dipole-moment surface, this good
agreement means that the superposition approximation used in Ref.~\onlinecite{Garberoglio_2021} for
the non-additive three-body polarizability is quite good, at least in the case of helium.  Since the
behavior of virial coefficients at high temperatures is more sensitive to the short-range details of
the potential and polarizability surfaces, we conclude that the superposition approximation is less
accurate at short range than it is at long range. This conclusion agrees with a detailed analysis of
the three-body polarizability surface performed in Ref.~\onlinecite{a3_2022}.

\begin{table*}
  \caption{Third dielectric virial coefficient $C_\varepsilon$ for
    ${}^3$He, its expanded uncertainty $U(C_\varepsilon)$, together with the various
    contributions from the pair potential
    ($C_\varepsilon^\mathrm{\therefore,2-body}$), the three-body potential and
    polarizability ($C_\varepsilon^{\mathrm{\therefore,diff-32}}$), the 
    dipole moment  ($C_\varepsilon^\mathrm{dip}$), and the exchange terms of
    Eqs.~(\ref{eq:Ceps1p}) and (\ref{eq:CepsCyc}). $U(C_\varepsilon)$ are
    expanded uncertainties at $k=2$ and include the statistical uncertainty
    of the calculation as well as the propagated uncertainties from
    potentials, polarizabilities, and dipole moment.}
  \begin{tabular}{d|d|d||d|d|d|d|d}
    \multicolumn{1}{c|}{Temperature} &
    \multicolumn{1}{c|}{$C_\varepsilon$} &
    \multicolumn{1}{c||}{$U(C_\varepsilon)$} &
    \multicolumn{1}{c|}{$C_\varepsilon^{\mathrm{\therefore,2-body}}$} &
    \multicolumn{1}{c|}{$C_\varepsilon^{\mathrm{\therefore,diff-32}}$} &    
    \multicolumn{1}{c|}{$C_\varepsilon^\mathrm{dip}$} &
    \multicolumn{1}{c|}{$C^{\cdot |}_\varepsilon$} &
    \multicolumn{1}{c}{$C^\triangle_\varepsilon$} \\    
    \multicolumn{1}{c|}{(K)} &
    \multicolumn{1}{c|}{(cm${}^9$/mol${}^3$)} & 
    \multicolumn{1}{c||}{(cm${}^9$/mol${}^3$)} &
    \multicolumn{1}{c|}{(cm${}^9$/mol${}^3$)} &
    \multicolumn{1}{c|}{(cm${}^9$/mol${}^3$)} &
    \multicolumn{1}{c|}{($10^{-5}$~cm${}^9$/mol${}^3$)} &    
    \multicolumn{1}{c|}{(cm${}^9$/mol${}^3$)} &   
    \multicolumn{1}{c}{(cm${}^9$/mol${}^3$)} \\
    \hline
1   &   -0.57   &   0.28    &   -0.268  &   -0.249  &   227.0   &   0.05425 &   -0.10561    \\
1.2 &   -0.50   &   0.19    &   -0.217  &   -0.236  &   181.4   &   0.01796 &   -0.06916    \\
1.4 &   -0.45   &   0.14    &   -0.182  &   -0.226  &   152.2   &   0.00412 &   -0.04802    \\
1.6 &   -0.42   &   0.10    &   -0.157  &   -0.228  &   128.1   &   -0.00651    &   -0.03131    \\
1.8 &   -0.39   &   0.09    &   -0.140  &   -0.223  &   112.4   &   -0.00785    &   -0.02180    \\
2   &   -0.37   &   0.08    &   -0.121  &   -0.225  &   96.6    &   -0.01053    &   -0.01505    \\
2.2 &   -0.33   &   0.07    &   -0.113  &   -0.198  &   85.7    &   -0.00890    &   -0.01035    \\
2.4 &   -0.33   &   0.06    &   -0.102  &   -0.209  &   77.1    &   -0.01071    &   -0.00730    \\
2.5 &   -0.31   &   0.06    &   -0.097  &   -0.199  &   72.6    &   -0.01014    &   -0.00611    \\
2.6 &   -0.31   &   0.06    &   -0.095  &   -0.209  &   69.7    &   -0.00656    &   -0.00518    \\
2.8 &   -0.29   &   0.05    &   -0.093  &   -0.191  &   64.2    &   -0.00474    &   -0.00372    \\
3   &   -0.29   &   0.05    &   -0.088  &   -0.197  &   58.8    &   -0.00557    &   -0.00264    \\
3.2 &   -0.27   &   0.05    &   -0.085  &   -0.184  &   54.3    &   -0.00469    &   -0.00190    \\
3.4 &   -0.27   &   0.04    &   -0.081  &   -0.183  &   50.7    &   -0.00289    &   -0.00135    \\
3.5 &   -0.27   &   0.04    &   -0.081  &   -0.186  &   49.4    &   -0.00305    &   -0.00115    \\
3.6 &   -0.26   &   0.04    &   -0.080  &   -0.173  &   47.6    &   -0.00302    &   -0.00099    \\
4   &   -0.26   &   0.04    &   -0.076  &   -0.181  &   42.6    &   -0.00233    &   -0.00052    \\
4.5 &   -0.26   &   0.03    &   -0.073  &   -0.181  &   37.4    &   -0.00119    &   -0.00025    \\
5   &   -0.24   &   0.03    &   -0.072  &   -0.167  &   33.3    &   -0.00069    &   -0.00012    \\
6   &   -0.23   &   0.03    &   -0.071  &   -0.163  &   27.7    &   -0.00026    &   -0.00003    \\
7   &   -0.24   &   0.02    &   -0.070  &   -0.165  &   23.8    &       &       \\
8   &   -0.23   &   0.02    &   -0.073  &   -0.157  &   21.0    &       &       \\
10  &   -0.23   &   0.02    &   -0.079  &   -0.152  &   17.3    &       &       \\
12  &   -0.24   &   0.02    &   -0.084  &   -0.152  &   14.9    &       &       \\
15  &   -0.24   &   0.02    &   -0.094  &   -0.147  &   12.7    &       &       \\
20  &   -0.25   &   0.02    &   -0.110  &   -0.145  &   10.6    &       &       \\
25  &   -0.26   &   0.02    &   -0.126  &   -0.136  &   9.4 &       &       \\
30  &   -0.28   &   0.02    &   -0.140  &   -0.137  &   8.7 &       &       \\
40  &   -0.30   &   0.02    &   -0.167  &   -0.137  &   7.9 &       &       \\
50  &   -0.32   &   0.02    &   -0.190  &   -0.135  &   7.6 &       &       \\
60  &   -0.35   &   0.02    &   -0.211  &   -0.136  &   7.4 &       &       \\
75  &   -0.36   &   0.02    &   -0.238  &   -0.127  &   7.4 &       &       \\
80  &   -0.37   &   0.02    &   -0.246  &   -0.125  &   7.4 &       &       \\
100 &   -0.40   &   0.02    &   -0.276  &   -0.125  &   7.6 &       &       \\
125 &   -0.43   &   0.02    &   -0.307  &   -0.122  &   7.9 &       &       \\
150 &   -0.45   &   0.02    &   -0.334  &   -0.118  &   8.2 &       &       \\
175 &   -0.47   &   0.02    &   -0.357  &   -0.115  &   8.6 &       &       \\
200 &   -0.49   &   0.03    &   -0.377  &   -0.112  &   9.0 &       &       \\
225 &   -0.50   &   0.03    &   -0.395  &   -0.110  &   9.5 &       &       \\
250 &   -0.52   &   0.03    &   -0.411  &   -0.110  &   9.9 &       &       \\
273.16  &   -0.53   &   0.03    &   -0.425  &   -0.106  &   10.3    &       &       \\
300 &   -0.54   &   0.03    &   -0.439  &   -0.102  &   10.7    &       &       \\
350 &   -0.56   &   0.03    &   -0.463  &   -0.101  &   11.5    &       &       \\
400 &   -0.58   &   0.03    &   -0.483  &   -0.098  &   12.3    &       &       \\
450 &   -0.60   &   0.03    &   -0.501  &   -0.097  &   13.1    &       &       \\
500 &   -0.61   &   0.03    &   -0.516  &   -0.096  &   13.9    &       &       \\
600 &   -0.62   &   0.03    &   -0.542  &   -0.083  &   15.4    &       &       \\
700 &   -0.64   &   0.04    &   -0.563  &   -0.074  &   16.9    &       &       \\
750 &   -0.64   &   0.04    &   -0.572  &   -0.071  &   17.6    &       &       \\
800 &   -0.64   &   0.04    &   -0.580  &   -0.063  &   18.3    &       &       \\
900 &   -0.65   &   0.04    &   -0.594  &   -0.057  &   19.8    &       &       \\
1000    &   -0.66   &   0.04    &   -0.606  &   -0.050  &   21.1    &       &       \\
1500    &   -0.67   &   0.04    &   -0.644  &   -0.026  &   27.7    &       &       \\
2000    &   -0.66   &   0.05    &   -0.662  &   0.001   &   33.9    &       &       \\
2500    &   -0.65   &   0.05    &   -0.670  &   0.019   &   39.8    &       &       \\
3000    &   -0.63   &   0.06    &   -0.672  &   0.045   &   45.5    &       &       
  \end{tabular}  
  \label{tab:He3}
\end{table*}

\begin{figure}
\includegraphics[width=0.95\linewidth]{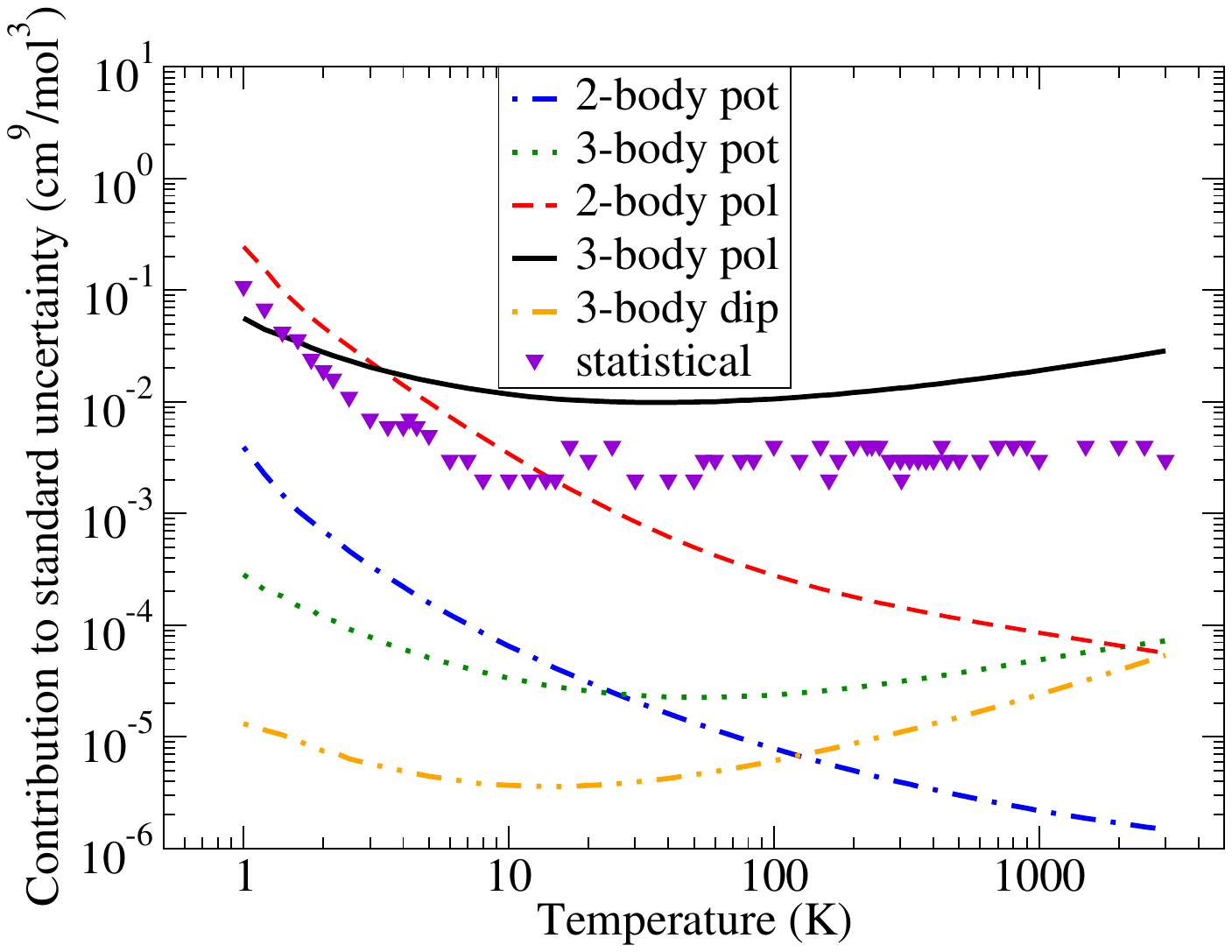}  
  \caption{The various contributions to the standard uncertainty of
    $C^\therefore_\varepsilon$ in the case of ${}^4$He. Solid line: the three-body
    polarizability. Dashed line: the pair polarizability. Dotted line:
    the three-body potential. Dot-dashed line: the pair
    potential. Dot-dot-dashed line: the three-body dipole. Triangles: statistical uncertainty from
    PIMC calculations.}
  \label{fig:unc}
\end{figure}

In our previous work, the use of the superposition approximation prevented a rigorous estimation of
the overall uncertainty of the PIMC calculations, in contrast to the results presented here.  As
might be expected by comparing the present values of $U(C_\varepsilon)$ to those of our previous
work,~\cite{Garberoglio_2021} the uncertainty budget of the third dielectric virial coefficient is
dominated by the uncertainty of the three-body polarizability surface.  In order to see that, a
breakdown of various components of the overall uncertainty is reported in Fig.~\ref{fig:unc} in the
case of the Boltzmann contribution $C^\therefore(T)$. For $T>10$~K, the largest contribution by far
to the overall propagated uncertainty comes from the uncertainty of the three-body polarizability,
followed by the propagated uncertainty from the two-body polarizability. The latter becomes the
dominant contribution for $T < 3.5$~K.
Figure~\ref{fig:unc} shows that we managed to keep the statistical uncertainty of the calculations
smaller than the largest contribution from the uncertainties propagated from potentials and
polarizabilities.
The contributions from the other three sources (the pair and
three-body potential energy, as well as the dipole moment) are found to be negligible overall, being
at least one order of magnitude smaller in the whole temperature range considered here.  The
individual uncertainty contributions presented in Fig.~\ref{fig:unc} are reported as Tables in the
Supplementary Material.

\begin{figure}
\includegraphics[width=0.95\linewidth]{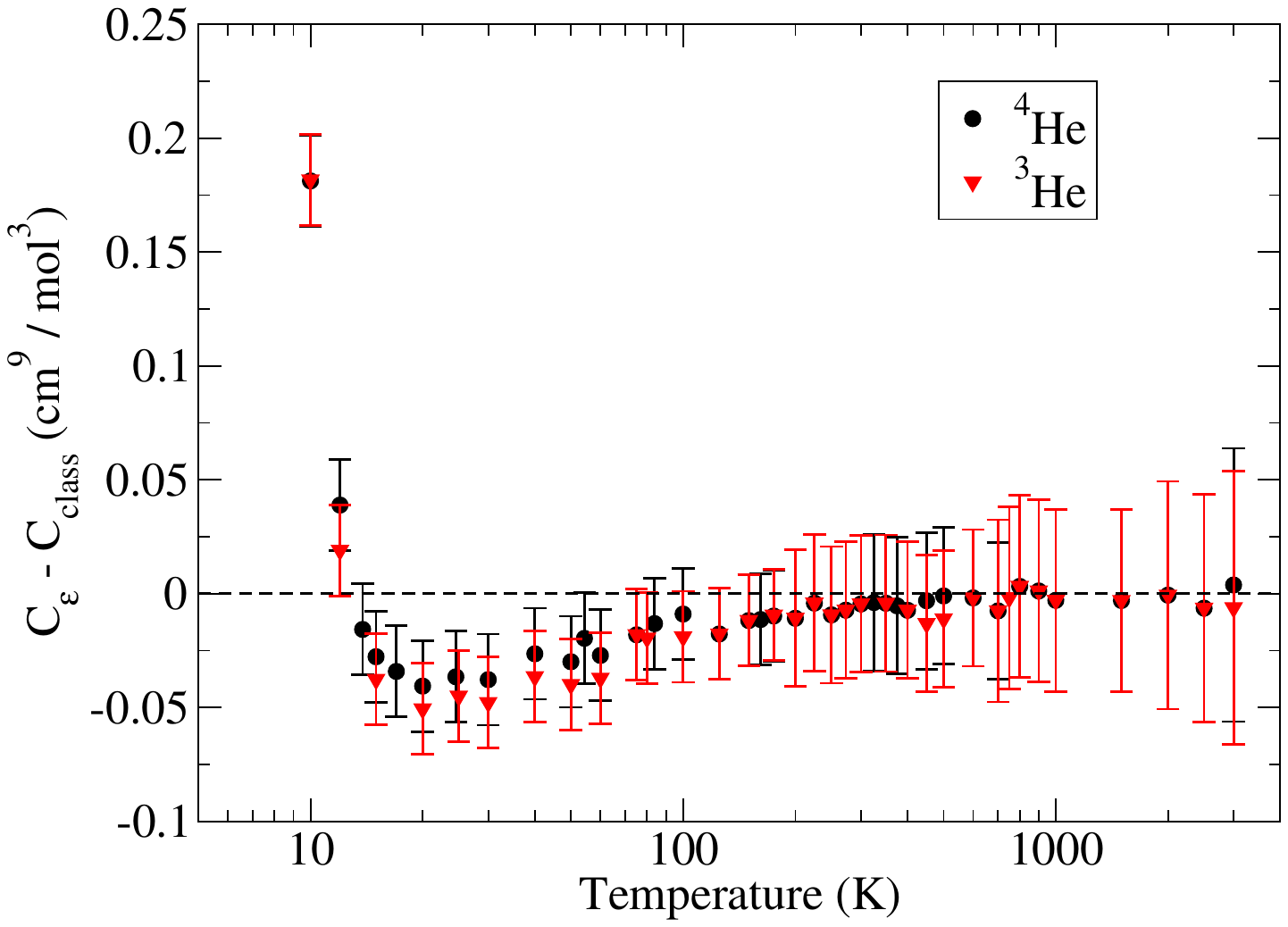}  
  \caption{The difference between the fully quantum calculation of $C_\varepsilon$ and the classical
    approximation. Circles: ${}^4$He, triangles: ${}^3$He. Error bars correspond to expanded $(k=2)$
    uncertainties.}
  \label{fig:classical}
\end{figure}

Finally, let us compare our fully quantum results with a classical
evaluation of the dielectric virial coefficient. As shown in
Fig.~\ref{fig:classical}, the classical approximation is quite good for
$T>200$~K. At lower temperatures, one can observe a systematic downward
shift, which becomes larger than the uncertainty of the calculation for
$T<50$~K. For still lower temperatures, quantum effects are responsible for
a significant upwards shift of the PIMC calculations.  As already noted
when comparing the quantum results for the two isotopes, the ${}^4$He and
${}^3$He results agree within their mutual uncertainties at all the
temperatures shown in Fig.~\ref{fig:classical}.

\subsection{Correlation for $C_\varepsilon$}

Values of $C_\varepsilon$ at temperatures not reported in Tables~\ref{tab:He4} and \ref{tab:He3} can
be obtained in principle by interpolating the tabulated values. However, given the relatively large
error bars, this procedure could produce spurious values.  Here we provide parameters for a smooth
correlation for both $C_\varepsilon$ and its $k=2$ uncertainty $U(C_\varepsilon)$.  For the former,
we note that the target accuracy of a fitting function should be given by only the statistical
uncertainty of the $C_\varepsilon$ calculation, since the other contributions to the uncertainty are
systematic.  We have fitted the results to the expansion
\begin{equation}
  C_\varepsilon(T) = \sum_{k=1}^6 \frac{a_k}{(T/T_0)^{b_k}},
  \label{eq:Ceps_correlation}
\end{equation}
with $T_0 = 30$~K, whereas for the expanded uncertainty we have used the representation
\begin{equation}
  U(C_\varepsilon)(T) = A_0 + \frac{A_1}{(T/{1~\mathrm{K}})^{c_1}} + A_2 (T/{1~\mathrm{K}})^{c_2},
  \label{eq:UCeps_correlation}
\end{equation}
whose parameters are reported in Table~\ref{tab:corr} for both ${}^4$He and ${}^3$He.  All the
values of $C_\varepsilon(T)$ computed using the correlation at the 59 temperatures considered in
this work (56 for ${}^3$He) fall within the expanded statistical uncertainty of the PIMC
calculation.

\begin{table}
  \caption{Parameters for the correlations (\ref{eq:Ceps_correlation}) and
    (\ref{eq:UCeps_correlation}) in the case of ${}^4$He and ${}^3$He. The $a_i$ and $A_i$ are given in
    cm${}^9$/mol${}^3$ while the $b_i$ and $c_i$ are dimensionless.}
  \begin{tabular}{c|d|d}
    Parameter&
    \multicolumn{1}{c|}{Value for ${}^4$He} &
    \multicolumn{1}{c}{Value for ${}^3$He} \\
    \hline
    $a_1$ & -2288.7466 & -363.45319 \\
    $a_2$ &  5191.1178 &  823.97628 \\
    $a_3$ & -4363.9948 & -902.48768 \\
    $a_4$ &  1461.3638 &  692.44074 \\
    $a_5$ & \multicolumn{1}{c|}{$-1.83960 \times 10^{-3}$}  & -250.75032 \\
    $b_1$ & \multicolumn{1}{c|}{9/20}  & \multicolumn{1}{c}{9/20}  \\
    $b_2$ & \multicolumn{1}{c|}{8/17}  & \multicolumn{1}{c}{1/2}  \\
    $b_3$ & \multicolumn{1}{c|}{1/2}   & \multicolumn{1}{c}{10/17}  \\
    $b_4$ & \multicolumn{1}{c|}{10/19} & \multicolumn{1}{c}{2/3}  \\
    $b_5$ & \multicolumn{1}{c|}{47/20} & \multicolumn{1}{c}{5/7} \\
    \hline
    $A_0$ & 0.025 & 0.02   \\
    $A_1$ & 0.56 & 0.25\\
    $c_1$   & 2.5  & 2\\
    $A_2$ &
    \multicolumn{1}{c|}{$1.3 \times 10^{-5}$} &
    \multicolumn{1}{c}{$2 \times 10^{-4}$} \\
    $c_2$   & 1 & \multicolumn{1}{c}{2/3} \\
  \end{tabular}
  \label{tab:corr}
\end{table}

\section{Comparison with experiment}
\begin{figure}
\includegraphics[width=0.95\linewidth]{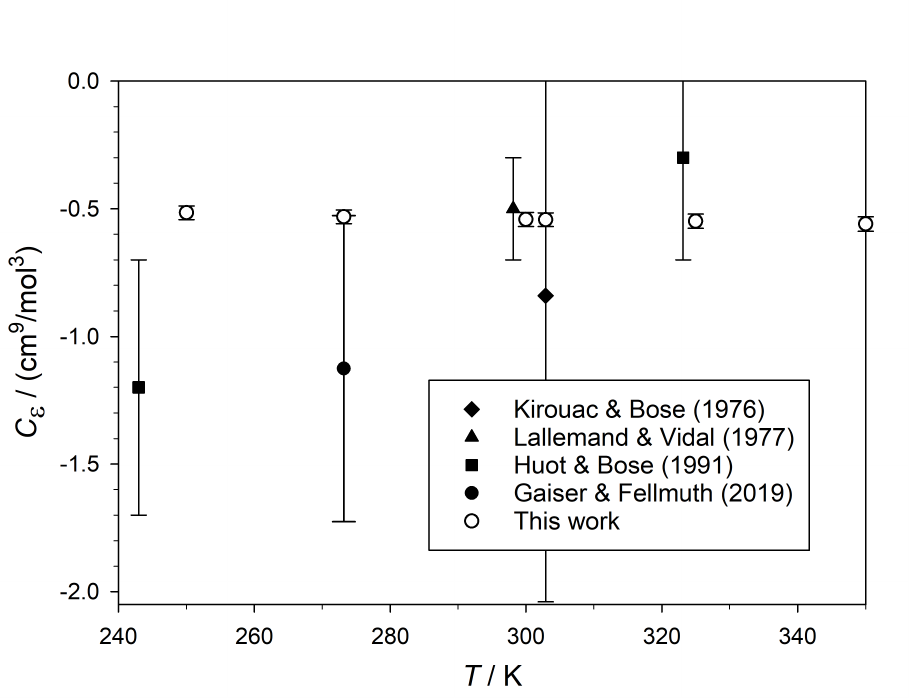}  
  \caption{Comparison of calculated values of the third dielectric virial coefficient $C_\varepsilon(T)$ for ${}^4$He with those derived from experiment\cite{Kirouac76,Lallemand77,Huot91,Gaiser_2019} near room temperature. Error bars for this work represent expanded ($k = 2$) uncertainties.}
\label{fig:He-high}  
\end{figure}

As a second-order correction, $C_\varepsilon$ is difficult to obtain from experiment.
A few measurements with large uncertainties exist for ${}^4$He; we are not aware of any experimental data for ${}^3$He.
Figure \ref{fig:He-high} shows the scattered data available for the third dielectric virial
coefficient of ${}^4$He near room temperature.  Our calculated results are the open circles; their
error bars, corresponding to expanded ($k = 2$) uncertainties, are similar to the size of the symbols.
For most of the experimental sources, the error bars are as given by the authors but the statistical meaning of their error interval was not stated.

The dielectric-constant gas thermometry measurements of Gaiser and Fellmuth\cite{Gaiser_2019} at
273.16~K did have a complete uncertainty budget; the corresponding point in Fig.~\ref{fig:He-high}
is plotted with $k = 2$ error bars.  Gaiser and Fellmuth reported a quantity that is a combination
of the second and third density virial coefficients and the second and third dielectric virial
coefficients.  Since $C_\varepsilon$ is the least accurately known of these four quantities, it can
be derived with an uncertainty essentially the same as that for the reported quantity.  To derive
$C_\varepsilon$, we used state-of-the-art values for the second\cite{u2_2020} and
third~\cite{Binosi_2024} density virial coefficients and for the second dielectric virial
coefficient,\cite{Garberoglio_2021} resulting in a value of \SI{-1.126}{cm^9.mol^{-3}} for
$C_\varepsilon$ with a $k = 2$ uncertainty of \SI{0.60}{cm^9.mol^{-3}}.  This is slightly less
negative than the value we derived and plotted in Ref.~\onlinecite{Garberoglio_2021} due to our use
of an improved value for the third density virial coefficient.~\cite{Binosi_2024} This datum is
consistent with our calculated result within its expanded uncertainty, but barely so.

\begin{figure}
\includegraphics[width=0.95\linewidth]{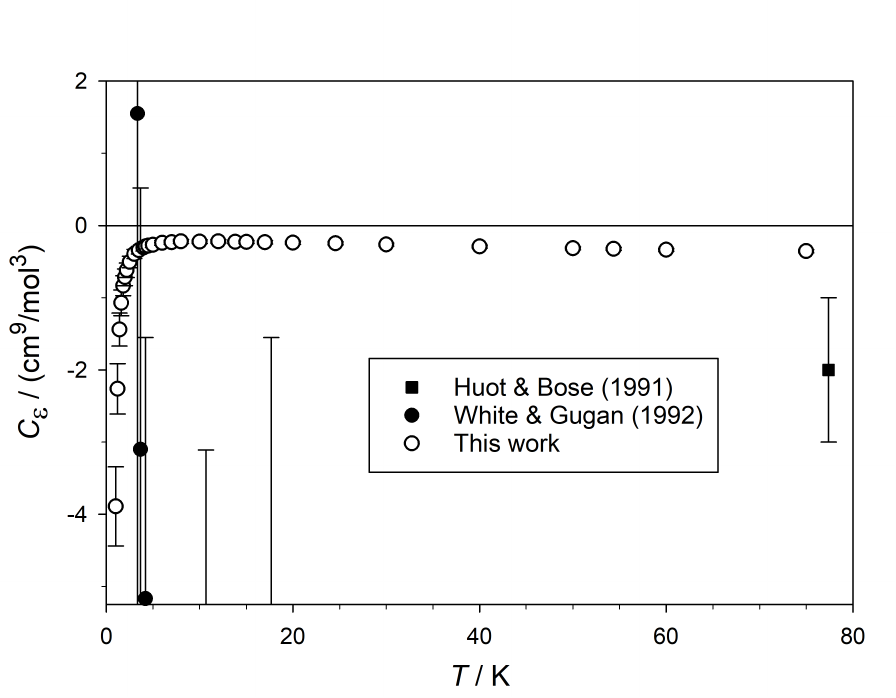}  
  \caption{Comparison of calculated values of the third dielectric virial coefficient $C_\varepsilon(T)$ for ${}^4$He with those derived from experiment\cite{Huot91,White_1992} at low temperatures. The two highest temperature points from Ref.~\onlinecite{White_1992} are below the bottom of the plot; only the tops of their error bars are visible.}
\label{fig:He-low}  
\end{figure}

Figure \ref{fig:He-low} shows the results at lower temperatures; the $k=2$ error bars for our calculated results are smaller than the size of the symbols above approximately 3~K.
Our result is in modest disagreement with the point from Huot and Bose\cite{Huot91} near 77~K, but with much smaller uncertainty.
At the lowest temperatures, the points from White and Gugan\cite{White_1992} have even larger uncertainties; they are at best in rough qualitative agreement with our calculations.

\begin{figure}
\includegraphics[width=0.95\linewidth]{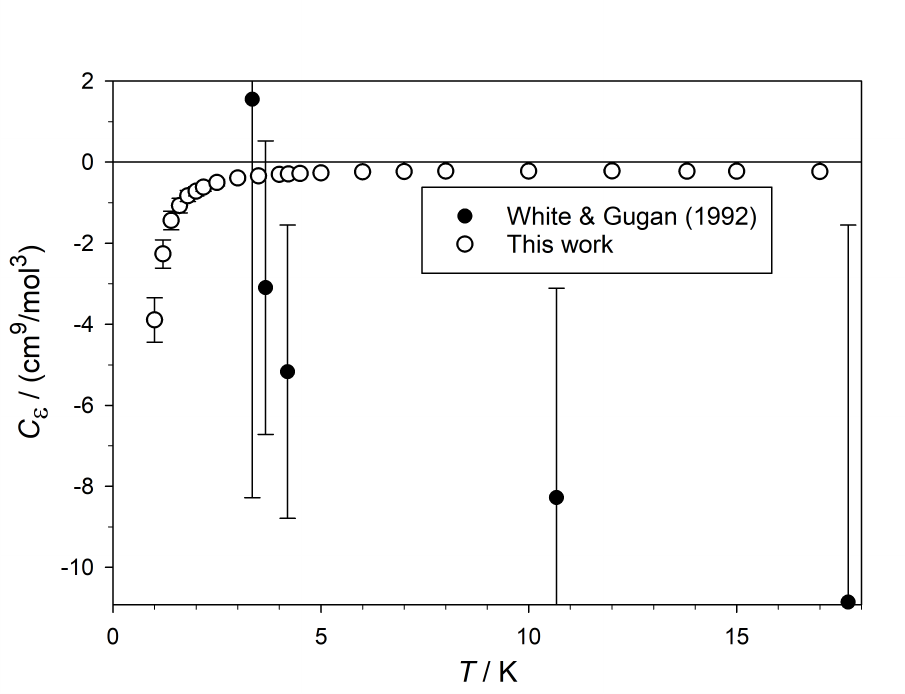}  
  \caption{Comparison of calculated values of the third dielectric virial coefficient $C_\varepsilon(T)$ for ${}^4$He with those from White and Gugan\cite{White_1992} below 18~K.}
\label{fig:He-lowest}  
\end{figure}

We examine the low-temperature data of White and Gugan\cite{White_1992} more closely in Fig.~\ref{fig:He-lowest}; on this scale the error bars for our calculations are only visible below approximately 2~K.
Figure~\ref{fig:He-lowest} makes clear that, if one were to infer a temperature dependence of $C_\varepsilon$ from the values reported by White and Gugan, the slope would have the wrong sign.
We note that even White and Gugan did not take their temperature dependence
too seriously; they ultimately recommended a single value of
$C_\varepsilon$ corresponding to ($-2.6 \pm 2.1$) \si{cm^9.mol^{-3}} to
represent their entire experimental range from 3~K to 18~K.  This value is
inconsistent with our results, but our points would be contained in their
large uncertainty range if their uncertainty estimate was larger by an additional
15\%. 
In Ref.~\onlinecite{Garberoglio20:Beps}, it was noted that the agreement
between highly accurate, rigorously calculated values of $B_\varepsilon$
and those reported by White and Gugan was strikingly good, especially
considering the challenging nature of their experiments at these low
temperatures.  It seems that this good agreement was limited to
$B_\varepsilon(T)$, and that White and Gugan did not derive meaningful
experimental values of $C_\varepsilon$. 

\section{Conclusions}

To support new developments in thermodynamic temperature and pressure metrology, this work presents
the first complete \textit{ab initio} calculation of the third dielectric virial coefficient,
$C_\varepsilon$, of helium. All physical effects are considered (including quantum statistical
effects that become important at low temperatures), and state-of-the-art first-principles surfaces
are used for the pair and three-body potentials, the pair and three-body polarizabilities, and the
three-body dipole moment.  Uncertainties from each of the surfaces, and from the path-integral Monte
Carlo calculations used to compute $C_\varepsilon$ with rigorous consideration of quantum effects,
are combined to yield a complete estimate for the uncertainty of $C_{\varepsilon}(T)$.

Since no previous \textit{ab initio} surface existed for the dipole moment of an assembly of three
helium atoms, a dipole-moment surface was developed in this work.  The \textit{ab initio}
calculations of the dipole moment accounted for electron correlation with the CC3 method and
employed doubly augmented QZ basis sets that had been developed previously.~\cite{cencek2012effects}
Some calculations with full correlation of all electrons and some with larger basis sets were performed
to assess uncertainty due to incompleteness of the calculations.  The results for the magnitude of
the three-body dipole moment were fitted to a function that was constrained to obey boundary
conditions known from theory.  At all temperatures considered, the contribution of the three-body
dipole term to $C_\varepsilon$ is roughly two orders of magnitude smaller than the expanded
uncertainty in $C_\varepsilon$, making it effectively negligible in this context. We could not have
drawn that conclusion with certainty without the present effort to calculate the contribution
rigorously.

Our results for $C_\varepsilon$ are generally consistent with previous work~\cite{Garberoglio_2021}
that employed approximations for the three-body polarizability and dipole-moment surfaces and that
used a now-obsolete surface for the three-body potential energy.  At temperatures above 300~K, our
new values of $C_\varepsilon$ are systematically higher (less negative), suggesting inaccuracy in
the short-range behavior of the superposition approximation for the three-body polarizability that
was used in Ref.~\onlinecite{Garberoglio_2021}.

At most temperatures, the uncertainty budget of $C_\varepsilon$ is dominated by the uncertainty of
the three-body polarizability surface (see Fig.~\ref{fig:unc}).  Since that surface was recently
developed with state-of-the-art methods,~\cite{a3_2022} improvement of that contribution in the near
future seems unlikely.  Below about 10~K, the nonadditive pair polarizability also makes a
significant contribution to the uncertainty budget, becoming the largest component below roughly
3.5~K.  The pair polarizability function is now more than ten years old,~\cite{Cencek11}
so improving it may be a worthwhile effort, especially if it is planned to use gas-based dielectric
or refractive measurements for temperature metrology below 10~K.

Our results also enable better refractivity-based metrology, the working equations of which employ a
virial expansion similar to Eq.~(\ref{eq:Vir:eps}) but with refractivity virial coefficients
$B_\mathrm{R}$ and $C_\mathrm{R}$.  At microwave frequencies, $C_\mathrm{R}$ is effectively
identical to the $C_\varepsilon$ derived here.  At optical and near-infrared frequencies, a
dispersion correction to $C_\varepsilon$ is needed for rigorous calculation of $C_\mathrm{R}$.  To
our knowledge, the information needed to make this correction has not been derived.  However, the
relatively small differences (less than 10\% and often less than 5\%) between $B_\varepsilon$ and
$B_\mathrm{R}$ for helium at optical frequencies~\cite{Garberoglio20:Beps}
suggests that, in the absence of other information, $C_{\varepsilon}(T)$ can provide a reasonable
approximation to $C_{\mathrm{R}}(T)$ in this regime.

\newpage

\section{Supplementary Material}

\begin{itemize}
\item Formulae for propagating the uncertainty of potentials, polarizabilities, and dipole moment to
  the various components of $C_\varepsilon$.
\item Table of the propagated uncertainties for ${}^4$He reported in Fig.~\ref{fig:unc}.
\item Plot illustrating the accuracy of the fit of the three-body dipole moment.
\item Raw \textit{ab initio} results CC3[d4Z] for all $734$ configurations considered in this work, and CC3[a3Z] and FCI[a3Z] results for configurations given in Table~\ref{tab:asympt}.
\item Optimized parameters of the fitting function~(\ref{eq:dip}).
\item FORTRAN program to compute the squared modulus of the dipole moment of a system of three He
  atoms.
\end{itemize}

\begin{acknowledgments}
B.J., G.G., and J.L. acknowledge support from {\em QuantumPascal} project 18SIB04, which has
received funding from the EMPIR programme co-financed by the Participating States and from the
European Union's Horizon 2020 research and innovation programme.  G.G. acknowledges CINECA (Award
No. IscraC-THIDIVI) under the ISCRA initiative for the availability of high-performance computing
resources and support. We gratefully acknowledge Poland's high-performance Infrastructure PLGrid
(HPC Centers: ACK Cyfronet AGH, PCSS, CI TASK, WCSS) for providing computer facilities and support
within computational grants PLG/2023/016599 and PLG/2024/017370.
\end{acknowledgments}

\section*{Author Declarations}
\subsection*{Conflict of Interest}
The authors have no conflicts to disclose.

\section*{Data Availability}
The data that support the findings of this study are available within the
article and its Supplementary Material.

\section{Supplementary Material}

\subsection{Formulae for the propagated uncertainties of $C_\varepsilon$}

\begin{widetext}
  
\subsubsection{Boltzmann component, $\therefore$}

These equations are already reported in the main text and are reproduced
here for completeness
\begin{eqnarray}
  \delta C_\varepsilon^{\therefore [u_2]}  &=&
  \frac{\pi \beta}{9} \int \left | \left\langle
  \sum_{i<j} \overline{\delta u_2}(\mbr_{ij}) \left[
    \left(\frac{\beta |\overline{\mbm_3^\therefore}|^2}{3} +
    \overline{A_3^\therefore} \right) \mathrm{e}^{-\beta
      \overline{V_3^\therefore}}
    - \overline{\alpha_\mathrm{iso}}(\mbr_{ij}) \e^{-\beta \overline{V_2^\therefore}(\mbr_{ij})}
    \right] -
  \right. \right. \nonumber \\
  & &
  \left. \left.
  6 \overline{\delta u_2}(\mbr_{12}) \e^{-\beta \overline{V_2^\therefore}(\mbr_{12})}
  \left(
  \e^{-\beta \overline{V_2^\therefore}(\mbr_{13})}
  (\overline{\alpha_\mathrm{iso}}(\mbr_{12}) +
  \overline{\alpha_\mathrm{iso}}(\mbr_{13}))
  -  \overline{\alpha_\mathrm{iso}}(\mbr_{12})
  \right)
  \right\rangle \right| \diff\mbr_2 \diff\mbr_3,
  \\
  \delta C_\varepsilon^{\therefore [u_3]}  &=&
  \frac{\pi \beta}{9} \int \left | \left\langle
  \overline{\delta u_3^\therefore} ~ 
  \left( \frac{\beta |\mbm_3^\therefore|^2}{3} + \overline{A_3^\therefore} \right)
  \e^{-\beta \overline{V_3^\therefore}}
  \right\rangle \right| \diff\mbr_2 \diff\mbr_3
  \\
  \delta C_\varepsilon^{\therefore [\alpha_2]}  &=&
  \frac{\pi}{9} \int \left | \left\langle
  \sum_{i<j} \overline{\delta \alpha_\mathrm{iso}}(\mbr_{ij}) \left(
  \e^{-\beta \overline{V_3^\therefore}} - \e^{-\beta \overline{V_2^\therefore}(\mbr_{ij})}
  \right) - \right. \right. \nonumber \\
  & & 
  \left. \left. 6 \left( \e^{-\beta \overline{V_2^\therefore}(\mbr_{12})} - 1 \right)
  \overline{\delta \alpha_\mathrm{iso}}(\mbr_{13}) \e^{-\beta \overline{V_2^\therefore}(\mbr_{13})}
  \right\rangle \right| \diff\mbr_2 \diff\mbr_3
  \\
  \delta C_\varepsilon^{\therefore [\alpha_3]} &=& \frac{\pi}{9}
  \int \left | \left\langle \overline{\delta \alpha_3^\therefore}
  \e^{-\beta \overline{V_3^\therefore}} \right\rangle \right| \diff\mbr_2
  \diff\mbr_3 
  \\  
  \delta C_\varepsilon^{\therefore [\mbm_3]}  &=& 
  \frac{2\pi \beta}{27} \int \left | \left\langle
  \overline{\delta\mbm_3^\therefore} \cdot \overline{\mbm_3^\therefore}
  \e^{-\beta \overline{V_3^\therefore}}  
  \right\rangle \right| \diff\mbr_2 \diff\mbr_3
\end{eqnarray}

\subsubsection{Odd component, $\cdot |$}

\begin{eqnarray}
  \delta C_\varepsilon^{{\cdot |} [u_2]}  &=&
  \frac{(-1)^{2I}}{2I+1}  \frac{\pi \beta \Lambda^3}{3 \cdot 2^{3/2}}
  \int \left | \left\langle  
  \sum_{i<j} \overline{\delta u_2^{\cdot |}}
  \left( \frac{\beta |\mbm_3^{\cdot |}|^2}{3} + 
  \overline{A_3^{\cdot |}} \right)
  \e^{-\beta V_3^{\cdot |}}
  - \overline{\delta u_2^|} \overline{\alpha_\mathrm{iso}^|} \e^{-\beta \overline{V_2^|}}  
  \right\rangle \right. \nonumber \\
  & & \left.
  -2 \left\langle \overline{\delta u_2^|} \e^{-\beta \overline{V_2^|}} \right\rangle
  \left\langle \overline{\alpha_\mathrm{iso}^:} \e^{-\beta \overline{V_2^:}} \right\rangle
  -2 \left\langle \e^{-\beta \overline{V_2^|}} \right\rangle
  \left\langle
  \overline{\delta u_2^:}
  \overline{\alpha_\mathrm{iso}^:}
  \e^{-\beta \overline{V_2^:}} \right\rangle
  \right . \nonumber \\
  & & \left .
  -2\langle \overline{\delta u_2^|} \overline{\alpha_\mathrm{iso}^|}
  \mathrm{e}^{-\beta \overline{V_2^|}} \rangle 
    \langle \mathrm{e}^{-\beta \overline{V_2^:}} -1 \rangle
  -2\langle \overline{\alpha_\mathrm{iso}^|} \mathrm{e}^{-\beta \overline{V_2^|}} \rangle
    \langle \overline{\delta u_2^:} \mathrm{e}^{-\beta \overline{V_2^:}} \rangle
  \right| \diff\mbr
  \label{eq:dCepsodd_u2}  \\  
  \delta C_\varepsilon^{{\cdot |} [u_3]}  &=&
  \frac{(-1)^{2I}}{2I+1}  \frac{\pi \beta \Lambda^3}{3 \cdot 2^{3/2}} \int \left | \left\langle
  \overline{\delta u_3^{\cdot |}} ~ 
  \left( \frac{\beta |\mbm_3^{\cdot |}|^2}{3} + 
  \overline{A_3^{\cdot |}} \right)
  \e^{-\beta \overline{V_3^{\cdot |}}}
  \right\rangle \right| \diff\mbr
  \label{eq:dCepsodd_u3}  \\
  \delta C_\varepsilon^{{\cdot |} [\alpha_2]} &=&
  \frac{(-1)^{2I}}{2I+1}  \frac{\pi \Lambda^3}{3 \cdot 2^{3/2}}
  \int \left | \left\langle
  \sum_{i<j} \overline{\delta\alpha_\mathrm{iso}^{\cdot |}}(\mbr_{ij}) \e^{-\beta V_3^{\cdot |}}
  - \overline{\delta\alpha_\mathrm{iso}^{|}} \e^{\-\beta \overline{V_2^|}}
  \right\rangle \right. \nonumber \\
  & & \left.
  -2 \left\langle \overline{\delta\alpha_\mathrm{iso}^{:}}(\mbr)
  \e^{-\beta \overline{V_2^:}(\mbr)}
  \right\rangle
  \left\langle \e^{-\beta \overline{V_2^|}}  \right\rangle  
  -2 \left\langle \overline{\delta\alpha_\mathrm{iso}^{|}} \e^{-\beta \overline{V_2^|}}\right\rangle
  \left\langle \e^{-\beta \overline{V_2^:}(\mbr)}-1 \right\rangle
  \right| \diff\mbr  
  \label{eq:dCepsodd_a2}\\  
  \delta C_\varepsilon^{{\cdot |} [\alpha_3]} &=&
  \frac{(-1)^{2I}}{2I+1}  \frac{\pi \Lambda^3}{3 \cdot 2^{3/2}}
  \int \left | \left\langle \overline{\delta \alpha_3^{\cdot |}}
  \e^{-\beta \overline{V_3^{\cdot |}}} \right\rangle \right| \diff\mbr
  \label{eq:dCepsodd_a3}\\
  \delta C_\varepsilon^{{\cdot |} [\mbm_3]}  &=&
  \frac{(-1)^{2I}}{2I+1} \frac{\pi \beta \Lambda^3}{9 \cdot 2^{3/2}}
  \int \left | \left\langle
  \overline{\delta\mbm_3^{\cdot |}} \cdot \overline{\mbm_3^{\cdot |}}
  \e^{-\beta \overline{V_3^{\cdot |}}}  
  \right\rangle \right| \diff\mbr
  \label{eq:dCepsodd_m}      
\end{eqnarray}

\subsubsection{Even component, $\triangle$}

\begin{eqnarray}
  \delta C_\varepsilon^{\triangle [u_2]}  &=&
  \frac{1}{(2I+1)^2}  \frac{2 \pi \beta \Lambda^3}{3}
  \left | \left\langle
  \frac{1}{3^{5/2}}
  \sum_{i<j} \overline{\delta u_2^|}(\mbr_{ij})
  \left( \frac{\beta |\mbm_3^\triangle|^2}{3} + 
  \overline{A_3^\triangle} \right)
  \e^{-\beta \overline{V_3^\triangle}}
  \right\rangle \right. \nonumber \\
  & & \left.
  -\frac{1}{4} 
  \left\langle
  \overline{\delta V_2^|}
  \mathrm{e}^{-\beta \overline{V_2^|}}
  \right\rangle
  \left\langle
  \overline{\alpha_\mathrm{iso}^|}
  \mathrm{e}^{-\beta \overline{V_2^|}}
  \right\rangle  
  -\frac{1}{4} 
  \left\langle
  \mathrm{e}^{-\beta \overline{V_2^|}}
  \right\rangle
  \left\langle
  \overline{\delta V_2^|}  
  \overline{\alpha_\mathrm{iso}^|}
  \mathrm{e}^{-\beta \overline{V_2^|}}
  \right\rangle  
  \right|
  \label{eq:dCepseven_u2}  \\  
  \delta C_\varepsilon^{\triangle [u_3]}  &=&
  \frac{1}{(2I+1)^2}  \frac{2 \pi \beta \Lambda^3}{3^{7/2}}
  \left | \left\langle
  \overline{\delta u_3^\triangle} ~ 
  \left( \frac{\beta |\mbm_3^\triangle|^2}{3} + 
  \overline{A_3^\triangle} \right)
  \e^{-\beta \overline{V_3^\triangle}}
  \right\rangle \right|
  \label{eq:dCepseven_u3}  \\  
  \delta C_\varepsilon^{\triangle [\alpha_2]} &=&
  \frac{1}{(2I+1)^2}  \frac{2 \pi \Lambda^6}{3}
  \left | \left\langle
  \frac{1}{3^{5/2}}
  \sum_{i<j}\overline{\delta \alpha_2^|}(\mbr_{ij})
  \e^{-\beta \overline{V_3^\triangle}} \right\rangle
  -\frac{1}{4}
  \left\langle \overline{\delta \alpha_\mathrm{iso}^|} \e^{-\beta
      \overline{V_2^|}} \right\rangle
  \left\langle \e^{-\beta \overline{V_2^|}} \right \rangle  
  \right| 
  \label{eq:dCepseven_a2}\\  
  \delta C_\varepsilon^{\triangle [\alpha_3]} &=&
  \frac{1}{(2I+1)^2}  \frac{2 \pi \Lambda^6}{3^{7/2}}
  \left | \left\langle \overline{\delta \alpha_3^\triangle}
  \e^{-\beta \overline{V_3^\triangle}} \right\rangle \right| 
  \label{eq:dCepseven_a3}\\
  \delta C_\varepsilon^{\triangle [\mbm_3]}  &=&
  \frac{1}{(2I+1)^2} \frac{2 \pi \beta \Lambda^6}{3^{9/2}}
  \left | \left\langle
  \overline{\delta\mbm_3^\triangle} \cdot \overline{\mbm_3^\triangle}
  \e^{-\beta \overline{V_3^\triangle}}  
  \right\rangle \right| 
  \label{eq:dCepseven_m}      
\end{eqnarray}

\subsection{Contributions to the uncertainty of $C_\varepsilon^\therefore$}

Table~\ref{tab:uncCeps} shows the data plotted in Figure 2 of the main
paper.

\begingroup
\begin{table}[h]
  \caption{Values of the components of the standard uncertainty of the Boltzmann
    part of $C_\varepsilon$ for ${}^4$He.}
  \begin{tabular}{c|c|c|c|c|c}
    Temperature & $u(V_2)$ & $u(V_3)$ & $u(\alpha_2)$ & $u(\alpha_3)$ &  $u(\mathbf{m}_3)$ \\
    (K) & (cm${}^9$/mol${}^3$) &(cm${}^9$/mol${}^3$) &(cm${}^9$/mol${}^3$) &(cm${}^9$/mol${}^3$) &(cm${}^9$/mol${}^3$) \\
    \hline
1	&	3.94E-03	&	2.84E-04	&	2.45E-01	&	5.61E-02	&	1.31E-05	\\
1.4	&	1.48E-03	&	1.81E-04	&	9.89E-02	&	3.89E-02	&	1.04E-05	\\
1.8	&	8.50E-04	&	1.39E-04	&	5.67E-02	&	3.05E-02	&	8.17E-06	\\
2	&	6.91E-04	&	1.13E-04	&	4.65E-02	&	2.78E-02	&	7.46E-06	\\
2.5	&	4.61E-04	&	9.21E-05	&	3.14E-02	&	2.34E-02	&	6.35E-06	\\
3	&	3.40E-04	&	7.83E-05	&	2.28E-02	&	2.04E-02	&	5.72E-06	\\
3.5	&	2.69E-04	&	6.77E-05	&	1.76E-02	&	1.86E-02	&	5.29E-06	\\
4	&	2.21E-04	&	6.05E-05	&	1.41E-02	&	1.72E-02	&	4.93E-06	\\
4.5	&	1.85E-04	&	5.70E-05	&	1.16E-02	&	1.61E-02	&	4.66E-06	\\
5	&	1.58E-04	&	5.08E-05	&	9.86E-03	&	1.53E-02	&	4.43E-06	\\
10	&	6.49E-05	&	3.35E-05	&	3.43E-03	&	1.17E-02	&	3.68E-06	\\
15	&	4.11E-05	&	2.81E-05	&	1.96E-03	&	1.06E-02	&	3.58E-06	\\
20	&	3.07E-05	&	2.57E-05	&	1.36E-03	&	1.02E-02	&	3.67E-06	\\
30	&	2.07E-05	&	2.35E-05	&	8.43E-04	&	9.89E-03	&	3.94E-06	\\
50	&	1.33E-05	&	2.26E-05	&	4.98E-04	&	9.94E-03	&	4.58E-06	\\
75	&	9.69E-06	&	2.30E-05	&	3.48E-04	&	1.03E-02	&	5.35E-06	\\
100	&	7.86E-06	&	2.36E-05	&	2.79E-04	&	1.06E-02	&	6.10E-06	\\
150	&	5.96E-06	&	2.56E-05	&	2.12E-04	&	1.14E-02	&	7.46E-06	\\
200	&	4.98E-06	&	2.75E-05	&	1.79E-04	&	1.21E-02	&	8.71E-06	\\
250	&	4.33E-06	&	2.94E-05	&	1.58E-04	&	1.27E-02	&	9.89E-06	\\
273.16	&	4.12E-06	&	3.03E-05	&	1.52E-04	&	1.30E-02	&	1.04E-05	\\
300	&	3.92E-06	&	3.12E-05	&	1.45E-04	&	1.33E-02	&	1.10E-05	\\
400	&	3.37E-06	&	3.45E-05	&	1.26E-04	&	1.43E-02	&	1.31E-05	\\
500	&	3.00E-06	&	3.73E-05	&	1.14E-04	&	1.53E-02	&	1.50E-05	\\
600	&	2.74E-06	&	4.00E-05	&	1.05E-04	&	1.61E-02	&	1.69E-05	\\
800	&	2.41E-06	&	4.48E-05	&	9.37E-05	&	1.77E-02	&	2.05E-05	\\
900	&	2.29E-06	&	4.69E-05	&	8.94E-05	&	1.83E-02	&	2.22E-05	\\
1000	&	2.18E-06	&	4.88E-05	&	8.56E-05	&	1.90E-02	&	2.39E-05	\\
2000	&	1.69E-06	&	6.32E-05	&	6.56E-05	&	2.44E-02	&	3.95E-05	\\
3000	&	1.48E-06	&	7.23E-05	&	5.62E-05	&	2.86E-02	&	5.37E-05	\\
\hline
  \end{tabular}
  \label{tab:uncCeps}
\end{table}  
\endgroup
  
\end{widetext}

\clearpage

\subsection{Fitting accuracy of the three-body dipole moment}

\begin{figure}[h!]
\includegraphics[width=0.95\linewidth]{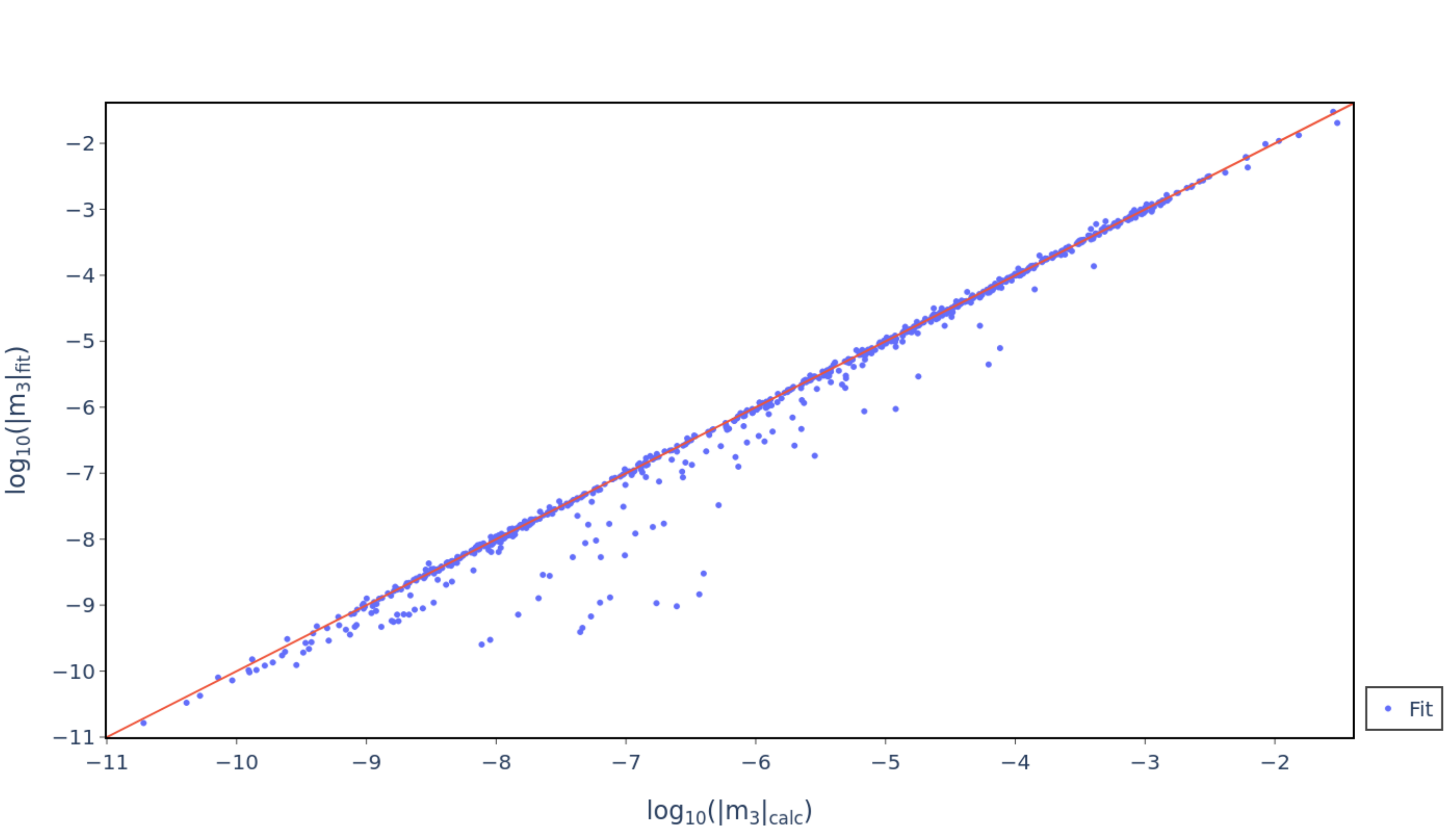}    
  \caption{Comparison of the fitted ($|\mathbf{m}_3|_\mathrm{fit}$) and \textit{ab initio}
    calculated ($|\mathbf{m}_3|_\mathrm{fit}$) lengths of the three-body dipole moments for helium
    for all $734$ geometries considered in the work. A logarithmic scale (base $10$) is used on both
    axes of the plot. A linear function $y=x$ is added in red for comparison.}
\label{fig:fitaccuracy}
\end{figure}

\end{document}